\documentclass[
    preprint,
    superscriptaddress, 
    prf,
    amsmath,amssymb,
    aps
]{revtex4-2}

\usepackage{graphicx} 
\usepackage{dcolumn}  
\usepackage{bm}       
\usepackage{hyperref} 
\usepackage{xcolor}   
\usepackage[scr=boondoxo]{mathalpha}

\hypersetup{
    colorlinks=true,
    linkcolor=blue,
    citecolor=blue,
    urlcolor=blue
}

\begin{document}

\title{Semi-analytical model for the rising sheet generated by droplet-pair impact}

\author{Shushan Hu}
\affiliation{
Institute of Thermal Science and Power Systems, School of Energy Engineering, Zhejiang University, Hangzhou, 310027, China 
}%


\author{Liwu Fan}
\email{liwufan@zju.edu.cn}
\affiliation{
Institute of Thermal Science and Power Systems, School of Energy Engineering, Zhejiang University, Hangzhou, 310027, China 
}%
\affiliation{State Key Laboratory of Clean Energy Utilization, Zhejiang University, Hangzhou 310027, China}

\author{Nan Hu}
\email{nh0529@princeton.edu} 
\affiliation{
Department of Mechanical and Aerospace Engineering, Princeton University, Princeton
}%
\date{\today}

\begin{abstract}
When two low-Ohnesorge-number drops impact a dry substrate simultaneously, their spreading lamellae collide and lift a free-standing vertical sheet. The sheet grows by inertial feeding from the spreading drops and is pulled back by capillary retraction at its rim. We develop a semi-analytical model for this rising sheet by extending the single-drop impact description of~\citet{Gordillo2019} to the two-drop geometry. The thin-film flow in the sheet is coupled at its base to the colliding lamellae and at its apex to a capillary-retarded rim. The sheet interior is then solved along ballistic characteristics in two stages: a lamella-fed stage, for which the velocity and thickness fields can be obtained in closed form, and a post-lamella stage, for which the inlet conditions are taken from simulations. The resulting framework gives the three-dimensional velocity and thickness fields and therefore the full sheet shape. On the centreline, the apex height and local thickness are obtained explicitly, showing that the different Weber-number exponents reported in the literature arise from a crossover rather than from a single universal scaling law. At sufficiently large Weber number, the apex pinches off. A linear Rayleigh--Plateau analysis, using the time-dependent jet diameter and deceleration predicted by the model, then bounds the maximum attainable height and closes the description of the pinch-off regime.
\end{abstract}

\maketitle

\section{Introduction}
The impact of liquid drops on solid surfaces is a classical problem in fluid mechanics that has been studied extensively for decades because of its rich interfacial dynamics and broad relevance to applications~\citep{Yarin2006,Josserand2016,Cheng2022}.For an isolated drop, the sequence of events following impact, including radial spreading, lamella formation and ejection, rim growth, and eventual retraction or breakup, is now reasonably well understood through a combination of experiments, numerical simulations, and reduced theory~\citep{CLANET2004,Yarin2006,Roisman2009,Eggers2010,Riboux2014,Wildeman2016,Gordillo2019,Sanjay2025}. In many practical situations, however, drops do not impact in isolation but as part of dense sprays, in which neighboring impacts interact on comparable spatial and temporal scales~\citep{Liang2016,Collision2017,Breitenbach2018}. Such interactions introduce collective dynamics that are absent from the single-drop problem. Yet, despite their practical importance, multi-drop impacts remain far less understood than the canonical case of a single impacting drop~\citep{Moreira2010}.

The simultaneous impact of two identical drops provides the simplest geometry in which such interaction effects appear. When the drops impact sufficiently close to each other, their spreading rims collide at the symmetry plane. The collision redirects part of the horizontal momentum upward and produces a free-standing sheet, hereafter referred to as the central rising sheet. The sheet is bounded at its upper edge by a retracting rim. This structure was described by \citet{Barnes1999}, who observed that it can destabilize and generate secondary droplets much larger than those produced by single-drop splashing. Subsequent studies have examined different aspects of this process. \citet{Roisman2002} proposed an early model for the sheet height, but without direct experimental validation. \citet{Ersoy2020} characterized the semilunar shape of the sheet. More recently, \citet{Goswami2023} measured the time-dependent sheet height, width, and thickness over a range of Weber numbers and drop spacings, and proposed an empirical scaling for the maximum height. \citet{Goswami2026} subsequently extended this characterization over nearly two decades in liquid viscosity, corresponding to Ohnesorge numbers from \(0.002\) to \(0.177\). They reported a morphological transition from well-defined semilunar sheets to short-lived liquid bumps, with capillary waves and rim corrugations progressively suppressed as viscosity increased. Beyond these experimental studies, \citet{Zhang2026} used three-dimensional direct numerical simulations to develop an energy-balance model for the maximum sheet height, in which the viscous dissipation was calibrated from the numerical data. Together, these studies have established the central rising sheet as the defining hydrodynamic feature of simultaneous two-drop impact. However, existing descriptions of its maximum height remain based on empirical scalings or numerical calibration. A predictive model for the unsteady sheet dynamics, linking the lamella collision to the sheet trajectory, thickness evolution, and rim motion, is still lacking.

Motivated by this gap, we develop a semi-analytical model for the unsteady central sheet. The model starts from the lamella description for single-drop spreading and extends it to the two-drop geometry. In \S\,\ref{sec:single}, we first summarize the elements of the single-drop lamella solution \citep{Gordillo2019} that provide the inlet conditions for the sheet. In \S,\ref{sec:3d}, we introduce the two-drop geometry, specify the matching conditions at the collision line and at the rim, and solve the three-dimensional thin-film problem by following the ballistic characteristics of fluid elements through the lamella-fed (L) and post-lamella (PL) stages. In the L stage, the explicit lamella solution from \citet{Gordillo2019} provides the boundary condition for the rising sheet, whereas in the PL stage we employ an empirical boundary condition extracted from fully numerical simulations. The resulting sheet profiles are then compared with experiments and simulations.
Since the apex height is the main observable, \S\,\ref{sec:sheet} reduces the three-dimensional solution to the centreline section. This reduction gives a closed-form expression for the height during the lamella-fed stage and a continuation into the post-lamella stage, which are tested against measured and simulated sheet. The model predicts an increasing height with Weber number, but in experiments the sheet height is ultimately limited by capillary breakup. We therefore close the description in \S\,\ref{pinch-off} by introducing a capillary cut-off: the time-dependent jet diameter and deceleration predicted by the model are used as inputs to a linear Rayleigh--Plateau analysis of the apex. Conclusions and limitations are given in \S\,\ref{sec:conclusions}.

\begin{figure}
  \centering
  \includegraphics[width=\textwidth]{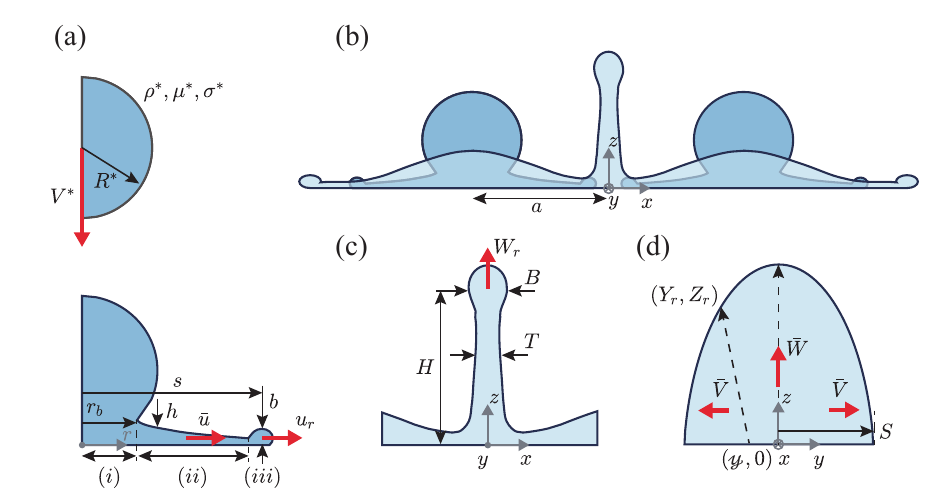}
  \caption{Schematic of droplet impact. (\textit{a}) Single droplet impact: (top) an undeformed spherical drop of radius $R^*$ and material properties $(\rho^*,\mu^*,\sigma^*)$ descending at velocity $V^*$ prior to impact, where $^*$ denotes quantities with dimensions; (bottom) post-impact spreading on the substrate, comprising (i) the undeformed drop region $r<\sqrt{3t}$, (ii) the lamella $\sqrt{3t}<r<s(t)$ with depth-averaged radial velocity $\bar{u}(r,t)$, and (iii) the rim of diameter $b(t)$ located at $s(t)$ and advancing at velocity $u_r(t)=\mathrm{d}s/\mathrm{d}t$. (\textit{b}) Simultaneous pair-drop impact at dimensionless half-spacing $a=\Delta x^*/(2R^*)$, with the two spreading lamellae meeting along the mid-plane and ejecting a central rising sheet. (\textit{c}) Cross-section of the central sheet at $y=0$, the plane to which the present analysis is restricted, of apex height $H(t)$ and local thickness $T(z,t)$, topped by a rim of diameter $B(t)$ rising at velocity $W_r(t)$. (\textit{d}) Front view of the central sheet (looking along $x$), illustrating its semilunar profile, the depth-averaged vertical velocity $\bar{W}(z,t)$ within the sheet, the lateral spreading velocity $\bar{V}$ of the surrounding lamellae whose divergence at $y=0$ produces the sink term $\phi_L T/t$ in the sheet continuity equation~(\ref{eq:sheet_cont}), and the collision front $S(t)$ marking the lateral extent of the meeting lamellae along $y$. Dashed lines trace the growth trajectory of a rim element from its origin $(\mathscr{y},0)$ on the collision front to its current position $(Y_r,Z_r)$ on the rim.}
  \label{fig:schematic}
\end{figure}

\section{Dynamics from a single impacting drop}
\label{sec:single}

Consider a droplet of dimensional radius $R^*$, density $\rho^*$, dynamic viscosity $\mu^*$, and surface tension $\sigma^*$ impacting a dry, flat substrate with normal velocity $V^*$. The time $t^*=0$ denotes the onset of impact. Quantities with an asterisk are dimensional, whereas all other variables are nondimensionalized using $R^*$ for lengths, $V^*$ for velocities, $R^*/V^*$ for times, and $\rho^*V^{*2}$ for stress. The governing dimensionless parameters are the Weber number $We=\rho^*V^{*2}R^*/\sigma^*$, the Reynolds number $Re=\rho^*V^*R^*/\mu^*$, and the Ohnesorge number $Oh=\mu^*/\sqrt{\rho^*R^*\sigma^*}=\sqrt{We}/Re$. We focus on the inertia-dominated regime relevant to millimetric drop impacts, for which $We\gg 1$ and $Re\gg 1$. The low-viscosity regime considered here corresponds to $10^{-3}\lesssim Oh\lesssim 10^{-2}$, consistent with the parameter range treated by~\citet{Gordillo2019}.

For the single-drop impact problem, we use cylindrical coordinates $(r,z)$ centered at the impact point. After contact, the liquid motion can be divided into three regions~\citep{Gordillo2019}, as shown in figure~\ref{fig:schematic}\textit{a}: the drop region, $0\leq r\leq r_b(t)$; the lamella region, $r_b(t)\leq r\leq s(t)$; and the rim region. The radius separating the drop and lamella regions is taken as $r_b(t)=\sqrt{3t}$. This expression was derived by~\citet{Riboux2014} from Wagner's linearized potential-flow theory~\citep{Wagner1932}. Although the derivation is asymptotically valid for $t\ll1$, comparisons with numerical lamella profiles show good agreement up to times of order unity $t=\mathscr{O}(1)$~\citep{Gordillo2019,Riboux2016}.

Within the lamella region, $\sqrt{3t}\le r\le s(t)$, the liquid spreads as a thin film with depth-averaged radial velocity $\bar u(r,t)$ and thickness $h(r,t)$. At the front, $r=s(t)$, surface tension collects the film into a toroidal rim. The governing equations for the lamella and rim, together with the solutions used here, are summarized in Appendix~\ref{sec:single_appendix} following~\citet{Gordillo2019}. These results provide the input for the pair-drop problem considered below.

The lamella model is based on three main approximations. First, pressure gradients within the lamella are neglected because the film is slender. Second, the outer lamella flow is treated as inviscid in the high-Reynolds-number limit. Third, the effect of the substrate enters through a thin viscous boundary layer, which provides the leading correction to the inviscid spreading flow. Accordingly, the depth-averaged velocity is expanded as
$\bar u=\bar u_0+Re^{-1/2}\bar u_1+O(Re^{-1})$, where $\bar u_0=r/t$ is the leading inviscid solution and $\bar u_1$ is the first correction due to boundary-layer friction at the wall. Once $\bar u(r,t)$ and $h(r,t)$ are known, the rim velocity $u_r(t)$ and rim radius $b(t)$ are obtained from the rim mass and momentum balances, using the lamella values $\bar u(s,t)$ and $h(s,t)$ at the front. For later convenience, we define $\phi(r,t)\equiv\bar u(r,t)/\bar u_0$, which measures the viscous attenuation of the radial lamella flow. 

We retain this first-order correction because it has an explicit analytical form Eq.~\eqref{eq:u1} and therefore allows the influence of wall-induced viscous attenuation on the subsequent sheet dynamics to be incorporated quantitatively. In particular, as shown in Eqs.~\eqref{eq:Hmax_scaling} and~\eqref{eq:We_eta}, this correction modifies the predicted scaling of the maximum sheet height, leading to a dependence distinct from the inviscid scaling $H_{\max}\sim We^{1/2}$.

\section{Three-dimensional dynamics of rising sheet}
\label{sec:3d}
\subsection{Geometry, notation and assumptions}
\label{Notation}
We now consider the simultaneous impact of two identical drops on a dry substrate, with dimensional center-to-center separation $\Delta x^*$, as shown in figure~\ref{fig:schematic}\textit{b}. Each drop has the same fluid properties and impact conditions as in \S\ref{sec:single}. The separation is measured by the dimensionless half-spacing
$a \equiv \Delta x^*/2R^*$.

We use a Cartesian coordinate system $(x,y,z)$ centered at the midpoint between the two impact points. The $z$ axis is normal to the substrate, the $x$ axis connects the two impact centers, and the $y$ axis lies along the collision line. With this convention, the central sheet rises in the $y$--$z$ plane and is thin in the $x$ direction. Throughout this section, lower-case symbols, such as $\bar u$, $b$, $h$, and $s$, denote lamella quantities inherited from the single-drop problem in \S\ref{sec:single}. Upper-case symbols are used for the central sheet: $\bar V(y,z,t)$ and $\bar W(y,z,t)$ are the depth-averaged velocity components in the $y$ and $z$ directions, $T(y,z,t)$ is the sheet thickness, $B(\mathscr{y},t)$ is the rim cross-sectional diameter, $H(t)$ is the apex height, and $S(t)$ is the collision front. These quantities are indicated in figure~\ref{fig:schematic}\textit{c,d}.

The model relies on several simplifying assumptions. We first assume that, before the two lamellae meet, the early spreading of each drop is unaffected by the presence of the other drop. The radial lamella velocity $\bar u$ and the rim motion $dr_b/dt$ are therefore taken directly from the single-drop problem in \S\ref{sec:single}. We also neglect the finite time required for the two drops to coalesce. When the two expanding rims first meet, the liquid contained in the individual rims is assumed to be incorporated instantaneously into the newly formed central sheet.

After collision, the two opposing radial fluxes meet at the symmetry plane. By symmetry, the velocity components normal to this plane, i.e. the $x$ components, are redirected upward by inertia, whereas the tangential components along the collision line, i.e. the $y$ components, are preserved~\citep{HassonPeck1964,Bremond2006,Roisman2002}. This redirection forms a free-standing sheet in the $y$--$z$ plane. Since the sheet is detached from the substrate except near its base, we neglect substrate friction within the sheet interior. Gravity is also neglected. For millimetric water drops impacting at $V^*\sim 2~\mathrm{m\,s^{-1}}$, the Froude number $Fr \equiv V^*/\sqrt{g^*R^*}\simeq 60$ is large, and the ratio of gravitational to inertial deceleration over the inertial--capillary time $t^*_{ic}=\sqrt{\rho^*{R^*}^3/\sigma^*}$ is
$g^*t^*_{ic}/V^*=We^{1/2}Fr^{-2}\lesssim 0.1$
over the range considered. Finally, after entering the free sheet, the flow is assumed to be inertia dominated. Each fluid parcel therefore follows a ballistic trajectory, with $D\mathbf{V}/Dt=\mathbf{0}$, while surface tension acts only through the bounding rim.

\begin{figure}
  \centering
  \includegraphics[width=\textwidth]{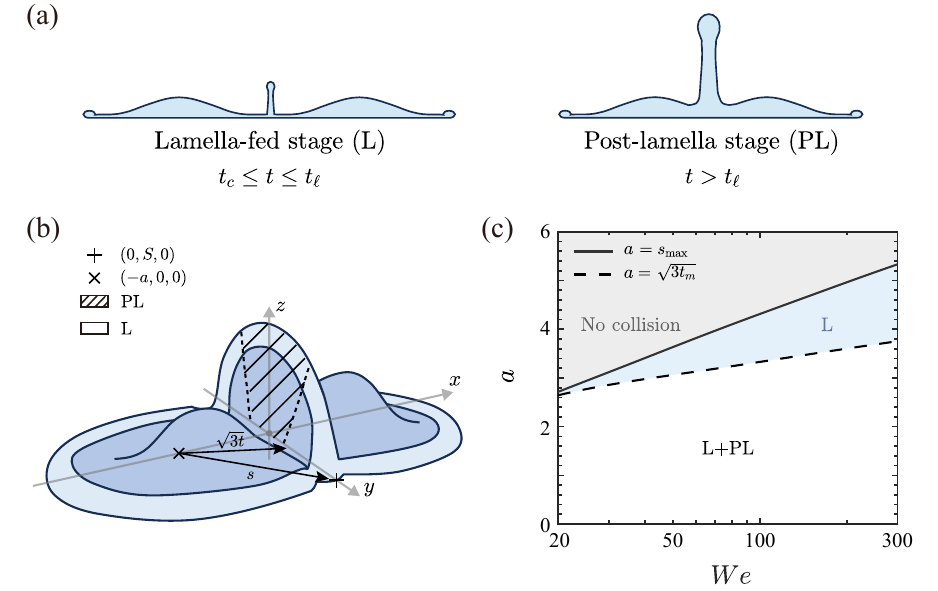}
  \caption{(\textit{a})~The two stages of the central sheet on the $y=0$ symmetry plane. In the lamella-fed stage (L, $t_c\leq t\leq t_\ell$) the parcel that reaches the apex is injected while the lamella still feeds the sheet, so the entire stall dynamics is contained in this stage; in the post-lamella stage (PL, $t>t_\ell$) the stall parcel is injected after feeding has ceased, and a two-stage integration is required. (\textit{b})~The full three-dimensional sheet at an instant $t$: points along the rim occupy different stages according to their lateral coordinate $\mathscr{y}$, separated by the boundary $\mathscr{r}=\sqrt{3t}$. (\textit{c})~Regime map in the $(We,a)$ plane for water : the solid line $a=s_{\max}$ bounds the region of no collision, and the dashed line $a=\sqrt{3t_{m}}$ separates the lamella-fed and post-lamella regimes.}
  \label{fig:phi_regime}
\end{figure}

\subsection{Dynamics and stages}
\label{sec:dynamics_stages}

Denote the maximum spreading radius of an isolated drop by $s_{\max}$. If
$a>s_{\max}$, the two drops do not interact. If $a<s_{\max}$, the two spreading
lamellae first meet at the midpoint $(x,y)=(0,0)$ at the collision onset time
$t_{c,0}$. For $t>t_{c,0}$, the collision line lies in the symmetry plane
$x=0$ and extends in the $y$ direction over $[-S(t),S(t)]$, where
$S(t)=\sqrt{s^2(t)-a^2}$. This quantity gives the lateral edge of the rising
sheet, as sketched in figure~\ref{fig:schematic}d. More generally, at a given
position $\mathscr{y}$ along the collision line, the local collision time
$t_c(\mathscr{y})$ is determined by
$s(t_c)=\sqrt{a^2+\mathscr{y}^2}$.

At the collision line, the two incoming lamellae carry equal and opposite
momentum in the $x$ direction. We assume that these normal components are
redirected into the vertical $z$ direction, while the tangential components
along $y$ are unchanged by the collision. At the point
$(0,\mathscr{y},0)$, with $|\mathscr{y}|\leq S$, the distance to either impact
centre is $\mathscr{r}=\sqrt{a^2+\mathscr{y}^2}$. The velocity of the lamella
coming from each parent drop can therefore be decomposed into components
$(\pm a\bar u(\mathscr{r},t)/\mathscr{r},
\mathscr{y}\bar u(\mathscr{r},t)/\mathscr{r})$, where the signs correspond to
the two drops centred at $(\mp a,0,0)$. After collision, the two contributions
give the inlet conditions for the central sheet:
$T(\mathscr{y},0,t)=2h(\mathscr{r},t)$,
$\bar W(\mathscr{y},0,t)=a\bar u(\mathscr{r},t)/\mathscr{r}$, and
$\bar V(\mathscr{y},0,t)=\mathscr{y}\bar u(\mathscr{r},t)/\mathscr{r}$.
Here $\bar u$ and $h$ are the single-drop lamella velocity and thickness
obtained in \S~\ref{sec:single}.

As the impact evolves, the boundary $r_b=\sqrt{3t}$ separating the drop and
lamella regions reaches the collision line first at $y=0$ and then moves
outward. In analogy with the local collision time, we
define the local lamella-ending time by
$t_{\ell}(\mathscr{y})=(a^2+\mathscr{y}^2)/3$, with
$t_{\ell,0}=a^2/3$ at $y=0$. Thus both $t_c$ and $t_{\ell}$ depend on
$\mathscr{y}$, and different portions of the sheet enter the subsequent stages
at different times. We refer to the interval
$t_c(\mathscr{y})\leq t\leq t_{\ell}(\mathscr{y})$ as the lamella-fed stage
(L), and to $t>t_{\ell}(\mathscr{y})$ as the post-lamella stage (PL).

In the PL stage, the thin lamella no longer supplies the sheet. The inlet is instead set by the bulk motion of the remaining liquid in the two drops, which continues to converge toward the symmetry plane after the lamella front has passed. This flow does not provide a closed-form inlet condition for the sheet, so
we prescribe the PL inlet from simulations. Specifically, we use
$T(\mathscr{y},0,t)=T^P(\mathscr{r},t)$,
$\bar W(\mathscr{y},0,t)=a\,\bar u_{PL}(\mathscr{r},t)/\mathscr{r}$, and
$\bar V(\mathscr{y},0,t)=\mathscr{y}\,\bar u_{PL}(\mathscr{r},t)/\mathscr{r}$.
Here, we empirically propose $T^P(\mathscr{r},t)$ is a polynomial in time,
{and $\bar u_{PL}(\mathscr{r},t)=C/t^{\zeta}$ with $\zeta=1.2$;} the constant $C$ is chosen to ensure continuity with the lamella-fed stage. These inlet forms are
empirical, in the same spirit as the thickness function $h_a$ used to close the single-drop theory~\citep{Riboux2016,Gordillo2019}. The distinction is that $h_a$ is used as part of a general single-drop description, whereas a given PL inlet fit is valid only over the range of conditions from which it was obtained.
The present fit covers all cases considered here; a substantially different parameter range would require a new calibration. Details of the extraction procedure and the range of validity are given in Appendix~\ref{app:inlet}.

For a fixed position $y=\mathscr{y}$, the local stage of the sheet is therefore
determined by the two times $t_c(\mathscr{y})$ and $t_{\ell}(\mathscr{y})$. The
sheet is in the L stage when
$t_c(\mathscr{y})\leq t\leq t_{\ell}(\mathscr{y})$, and in the PL stage when
$t>t_{\ell}(\mathscr{y})$, as shown in figure~\ref{fig:phi_regime}a. At a
given time $t>t_{\ell,0}$, different parts of the same sheet can therefore be
in different stages, as illustrated in figure~\ref{fig:phi_regime}b. The inlet
conditions for the sheet can then be written compactly as
\begin{equation}
\label{BC_Sheet}
\begin{aligned}
T(\mathscr{y},0,t) \equiv T_0 &=
\begin{cases}
2h(\mathscr{r},t),
& t_c(\mathscr{y}) \leq t \leq t_{\ell}(\mathscr{y}), \\[6pt]
T^P(\mathscr{r},t),
& t > t_{\ell}(\mathscr{y}),
\end{cases} \\[8pt]
\bar W(\mathscr{y},0,t) \equiv \bar W_0 &=
\begin{cases}
\dfrac{a}{\mathscr{r}}\,\bar u(\mathscr{r},t),
& t_c(\mathscr{y}) \leq t \leq t_{\ell}(\mathscr{y}), \\[8pt]
\dfrac{a}{\mathscr{r}}\,\bar u_{PL}(\mathscr{r},t),
& t > t_{\ell}(\mathscr{y}),
\end{cases} \\[8pt]
\bar V(\mathscr{y},0,t) \equiv \bar V_0 &=
\begin{cases}
\dfrac{\mathscr{y}}{\mathscr{r}}\,\bar u(\mathscr{r},t),
& t_c(\mathscr{y}) \leq t \leq t_{\ell}(\mathscr{y}), \\[8pt]
\dfrac{\mathscr{y}}{\mathscr{r}}\,\bar u_{PL}(\mathscr{r},t),
& t > t_{\ell}(\mathscr{y}).
\end{cases}
\end{aligned}
\end{equation}
It should be noted that $h$ and $\bar u$ are analytically obtained as shown in Appendix~\ref{sec:single_appendix}, while $T^p$ and $\bar u _{PL}$ are empirically obtained as shown in Appendix~\ref{app:inlet}.

Since the present study focuses on the rise of the central sheet, we define $t_m$ as the time at which the centreline height reaches its maximum value $H_m$. Comparing $t_m$ with $t_{\ell,0}$ divides the $a$--$We$ phase diagram into three regimes: no collision, L only, and L+PL, as shown in figure~\ref{fig:phi_regime}c. The L-only regime occupies only a narrow part of the phase diagram, so the remainder of the paper focuses mainly on the sheet dynamics in the L+PL regime.

\subsection{Semi-analytical model}
\label{Approximate_model}
\subsubsection{Sheet domain}
\label{sheet_domain}
We first describe the velocity and thickness fields in the sheet interior, away from the bounding rim. The sheet is fed through the plane \(z=0\), where the values of \(T\), \(\bar V\), and \(\bar W\) are prescribed by the stage-dependent conditions \eqref{BC_Sheet}. After entering the free sheet, the fluid is assumed to experience neither viscous friction nor gravitational deceleration. Its velocity therefore remains constant along material paths, while the thickness evolves by mass conservation,
\begin{equation}
  \frac{D\bar{\mathbf V}}{Dt}=\mathbf 0,
  \qquad
  \frac{\partial T}{\partial t}
  +\frac{\partial(\bar V T)}{\partial y}
  +\frac{\partial(\bar W T)}{\partial z}=0 .
  \label{eq:governing1}
\end{equation}
Here
$\frac{D}{Dt}
  =
  \frac{\partial}{\partial t}
  +\bar V\frac{\partial}{\partial y}
  +\bar W\frac{\partial}{\partial z}$
and
$\bar{\mathbf V}
  =
  \bar V\,\mathbf e_y+\bar W\,\mathbf e_z$.
Consider a fluid element injected at time \(\tau\) from
\((\mathscr y,0)\), with inlet velocity \((\bar V_0,\bar W_0)\). Since the velocity is constant along its trajectory, its position at a later time \(t\) is
\begin{equation}
  y=\mathscr y+\bar V_0(t-\tau),
  \qquad
  z=\bar W_0(t-\tau).
  \label{eq:ballistic_map_general}
\end{equation}
Expanding the mass-conservation equation
then gives the thickness evolution along the same characteristic,
\begin{equation}
  \frac{DT}{Dt}
  =
  -T\left(
  \frac{\partial \bar V}{\partial y}
  +
  \frac{\partial \bar W}{\partial z}
  \right).
  \label{eq:T_characteristic_general}
\end{equation}

{
We now apply this construction to the L stage. On the inlet plane $z=0$ the conditions are those of~\eqref{BC_Sheet}, with the lamella velocity written as $\bar u(\mathscr r,t)=\phi(\mathscr r,t)\,\mathscr r/t$, so that
\begin{equation}
  \bar V_0=\phi\,\frac{\mathscr y}{t},\qquad
  \bar W_0=\phi\,\frac{a}{t},\qquad
  T_0=2h(\mathscr r,t).
  \label{eq:inject_3d}
\end{equation}
Applying the construction~\eqref{eq:ballistic_map_general} and~\eqref{eq:T_characteristic_general} with this inlet, the attenuation carried along each characteristic is its value at the origin, $\phi_L\equiv\phi(\mathscr r,\tau)$, and is therefore fixed once $\tau$ is fixed, since the velocity is conserved on material paths~\eqref{eq:governing1}. The time dependence of $\phi$ is retained: it enters the final fields through $\phi_L=\phi(\mathscr r,\tau(z,t))$, which varies from one characteristic to the next.} At the inlet, the ratio of the vertical to horizontal velocity components is \(\bar W_0/\bar V_0=a/\mathscr y\). Since both components remain constant along a characteristic, this ratio is preserved as the fluid element moves through the sheet. The characteristic starting from \((\mathscr y,0)\) therefore satisfies
\begin{equation}
  z=(y-\mathscr y)a/\mathscr y,
  \qquad
  \mathscr y=\frac{ay}{a+z}.
  \label{eq:traj}
\end{equation}
The corresponding injection time is obtained from the vertical displacement,
which gives
\begin{equation}
  \tau=\frac{\phi_L a}{z+\phi_L a}\,t .
  \label{eq:tau_L}
\end{equation}
Because the velocity is constant along each characteristic,
\(\bar{\mathbf V}(y,z,t)=\bar{\mathbf V}(\mathscr y,0,\tau)\). Substituting
\eqref{eq:traj} and \eqref{eq:tau_L} into the inlet velocity then gives the
L-stage velocity field
\begin{equation}
  \bar W(y,z,t)=\frac{z+\phi_L a}{t},
  \qquad
  \bar V(y,z,t)=
  \frac{(z+\phi_L a)y}{(a+z)t}.
  \label{eq:vel_field_3d}
\end{equation}
The thickness is then obtained from \eqref{eq:T_characteristic_general}.  From
\eqref{eq:vel_field_3d}, we have
\(\partial\bar W/\partial z=1/t\) and
\(\partial\bar V/\partial y=(z+\phi_L a)/[(a+z)t]\).  Hence, along a
characteristic,
\(\mathrm DT/\mathrm Dt=-T\{1/t+(z+\phi_L a)/[(a+z)t]\}\).  On the same
characteristic, \(Dz/Dt=\bar W=(z+\phi_L a)/t\), so
\(D\log(z+\phi_L a)/Dt=1/t\) and
\(D\log(a+z)/Dt=(z+\phi_L a)/[(a+z)t]\).  Combining these relations gives
\begin{equation}
  \frac{D}{Dt}
  \left[
  T(z+\phi_L a)(a+z)
  \right]=0 .
  \label{eq:T_conserved}
\end{equation}
The bracketed quantity is therefore constant along each characteristic.  At the
injection point, \(z=0\), \(T=2h(\mathscr r,\tau)\), and
\((z+\phi_L a)(a+z)=\phi_L a^2\).  Thus
\begin{equation}
  T(y,z,t)
  =
  2h(\mathscr r,\tau)
  \frac{\phi_L a^2}{(z+\phi_L a)(a+z)} .
  \label{eq:T_field}
\end{equation}

The PL stage is treated in the same way, but with different inlet boundary conditions at
\(z=0\). In this stage,
\begin{equation}
    T_0=T^P(\mathscr r,t),
  \qquad
  \bar W_0=\frac{a}{r}\bar u_{PL}(\mathscr r,t),
  \qquad
  \bar V_0=\frac{\mathscr y}{r}\bar u_{PL}(\mathscr r,t).
\end{equation}
In this case the relation between \((y,z,t)\) and the injection variables
\((\mathscr y,\tau)\) is no longer algebraic.  We therefore solve the inverse
characteristic problem numerically.  Once \((\mathscr y,\tau)\) has been found,
the thickness is obtained by integrating
\eqref{eq:T_characteristic_general} along the same characteristic.

\subsubsection{Rim domain}
\label{rim_domain}

The sheet solution provides the velocity and thickness incident on the rim. We
describe the rim by following material elements labelled by their source
coordinate $\mathscr y\in[-S,S]$. The element labelled by $\mathscr y$ has
position $(Y_r,Z_r)$, velocity $\mathbf V_r=(V_r,W_r)$, and cross-sectional
diameter $B(\mathscr y,t)$. The derivative $\mathrm d/\mathrm dt$ is taken at
fixed $\mathscr y$.

For a given $\mathscr y$, the direction of the incoming sheet flow is fixed by
the impact geometry. We define
\[
\hat{\mathbf e}(\mathscr y)
=
\frac{\mathscr y}{\mathscr r}\,\mathbf e_y
+
\frac{a}{\mathscr r}\,\mathbf e_z,
\qquad
\mathscr r=\sqrt{a^2+\mathscr y^2}.
\]
The incident sheet velocity $\bar{\mathbf V}=(\bar V,\bar W)$ and thickness
$T$ are evaluated at the current rim position $(Y_r,Z_r,t)$. The local flux
incorporated into the rim element is then the sheet thickness multiplied by the
relative velocity projected onto the incoming direction,
\begin{equation}
  \dot m
  =
  T\left[
  (\bar V-V_r)\frac{\mathscr y}{\mathscr r}
  +
  (\bar W-W_r)\frac{a}{\mathscr r}
  \right].
  \label{eq:rim_flux}
\end{equation}
Here $\dot m$ is defined for the rim element labelled by $\mathscr y$.
Mass conservation gives
\begin{equation}
  \frac{\pi}{4}\frac{\mathrm{d}B^{2}}{\mathrm{d}t}=\dot m,
  \label{eq:rim_mass_3d}
\end{equation}
where $\pi B^2/4$ is the cross-sectional area of the rim. Momentum conservation
for the same element gives
\begin{equation}
  \frac{\pi B^{2}}{4}\frac{\mathrm{d}\mathbf V_r}{\mathrm{d}t}
  =
  \dot m\,(\bar{\mathbf V}-\mathbf V_r)
  -
  \frac{2}{We}\hat{\mathbf e}.
  \label{eq:rim_mom_3d}
\end{equation}
The first term on the right-hand side is the momentum brought into the rim by
the incorporated sheet flux. The second term is the capillary retraction force;
its magnitude is $2/We$ because the rim bounds a free sheet with two liquid--air
interfaces. These equations correspond to the single-drop rim balance
\eqref{eq:rim_single} with the free-rim geometric and capillary coefficients
$\alpha=1$ and $\beta=1$. The rim position is advanced according to
$\mathrm{d}(Y_r,Z_r)/\mathrm{d}t=\mathbf V_r$. In components,
\eqref{eq:rim_mom_3d} becomes
\begin{equation}
  \frac{\pi B^{2}}{4}\frac{\mathrm{d}V_r}{\mathrm{d}t}
  =
  \dot m\,(\bar V-V_r)
  -
  \frac{2}{We}\frac{\mathscr y}{\mathscr r},
  \qquad
  \frac{\pi B^{2}}{4}\frac{\mathrm{d}W_r}{\mathrm{d}t}
  =
  \dot m\,(\bar W-W_r)
  -
  \frac{2}{We}\frac{a}{\mathscr r}.
  \label{eq:rim_mom_components}
\end{equation}

The initial conditions are imposed when each rim element is created at the
collision line, at $t=t_c(\mathscr y)$. At this time,
\begin{equation}
  Y_r(\mathscr y, t_c) = \mathscr y, \qquad Z_r(\mathscr y, t_c) = 0 .
  \label{eq:rim_init_pos}
\end{equation}
The newly formed rim element contains the local sheet feed and the two
single-drop rims that collide at the same location. Its initial speed is taken
as the mass-weighted average
\begin{equation}
  V_{\rm avg}(\mathscr y) = \frac{m_s\,\bar u(\mathscr r,t_c) + 2\,m_r\,u_r(t_c)}{m_s + 2\,m_r},
  \label{eq:Vavg_init}
\end{equation}
where $m_s = \pi\,T(\mathscr y,0,t_c)^2/4$ and
$m_r = \pi\,b(t_c)^2/8$ are the effective cross-sectional areas associated with
the sheet feed and with each incoming single-drop rim, respectively. Projecting
this speed onto the incoming direction gives the initial rim velocity,
\begin{equation}
  V_r(\mathscr y,t_c) = V_{\rm avg}\frac{\mathscr y}{\mathscr r}, \qquad
  W_r(\mathscr y,t_c) = V_{\rm avg}\frac{a}{\mathscr r}, \qquad
  \frac{\pi}{4}B(\mathscr y,t_c)^2 = m_s + 2\,m_r.
  \label{eq:rim_init_vel}
\end{equation}

Equations~\eqref{eq:rim_flux}--\eqref{eq:rim_init_vel}, driven by the sheet
interior solution, form a closed system for the motion and growth of the rim.
Integrating this system for each source coordinate
$\mathscr y\in[-S,S]$ gives the full side-view profile of the rising sheet in
the $(y,z)$ plane, without any additional adjustable parameter in the rim model.

\subsubsection{Comparison with experiment and simulation}
\begin{figure}
  \centering
  \includegraphics[width=\textwidth]{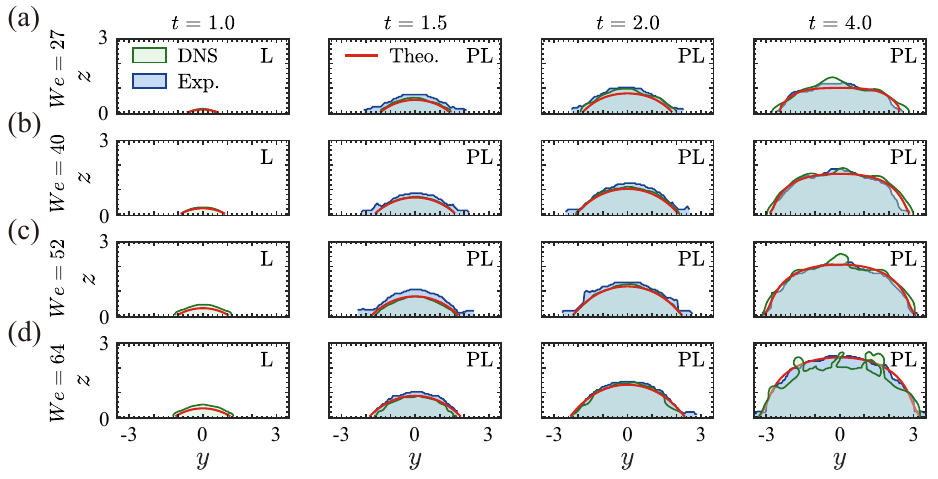}
  \caption{Side-view profiles of the central sheet at four Weber numbers $We\in\{27,40,52,64\}$ (rows) and four post-impact times $t\in\{1.0,1.5,2.0,4.0\}$ (columns), at fixed $a=1.80$ and $Oh=2.9\times10^{-3}$. Red lines are the present theory, obtained by integrating the rim system~\eqref{eq:rim_mom_components}; blue lines are the experiments of \citet{Goswami2023} and green lines the present simulations.}
  \label{fig:exp-vs-model}
\end{figure}
{
The sheet fields of \S\ref{sheet_domain} and the rim balances of \S\ref{rim_domain} together close the model. The sheet solution of \S\ref{sheet_domain} is obtained separately and enters the rim problem as a known field, so the variables advanced in time are the rim position, velocity and cross-section, governed by~\eqref{eq:rim_mass_3d}--\eqref{eq:rim_mom_components} with $\mathrm d(Y_r,Z_r)/\mathrm dt=\mathbf V_r$; integrating them fixes the sheet profile and the apex height $H(t)$. We discretize the source coordinate $\mathscr y$ into a fixed set of values along the collision line and integrate the rim system for each element, the value $\mathscr y=0$ giving the apex. An element is created at its collision time $t_c(\mathscr y)$ with the initial state~\eqref{eq:rim_init_pos} and~\eqref{eq:rim_init_vel}. At each step the sheet solution of \S\ref{sheet_domain}, driven by the stage-dependent inlet~\eqref{BC_Sheet}, is evaluated at each element's current position $(Y_r,Z_r,t)$ to supply the incident velocity $\bar{\mathbf V}$ and thickness $T$, each element drawing on the lamella-fed or post-lamella branch according to its own $t_\ell(\mathscr y)$. The incident $\bar{\mathbf V}$ and $T$ drive the flux~\eqref{eq:rim_flux} and the balances~\eqref{eq:rim_mass_3d}--\eqref{eq:rim_mom_components}, which are integrated by an explicit scheme at a fixed step $\Delta t=10^{-3}$. Halving the step $\Delta t$ and doubling the number of elements in $\mathscr y$ leave the profiles below unchanged; the apex height is read from the $\mathscr y=0$ element and the side-view profile from the element positions in the $(y,z)$ plane.

We now compare the predicted profiles with experiments and with direct numerical simulations (DNS). The DNS, described in Appendix~\ref{sec:dns_appendix}, provide an independent reference unaffected by the optical occlusion of the experimental side view, in which the central sheet can be partially hidden by the surrounding drop bulk.
}
We first vary the Weber number at fixed half-spacing. Figure~\ref{fig:exp-vs-model} compares the predicted profiles with the experiments of~\citet{Goswami2023} and with our simulations for four Weber numbers and four post-impact times at $a=1.80$. The semi-analytical solution captures the overall sheet shape in the $y$--$z$ plane, the growth of both the apex height and the basal footprint, and the increasing steepness of the dome as $We$ increases. The agreement is close at early times for all four Weber numbers. At the latest time, $t=4.0$, the high-$We$ sheets have become unstable and pinch off in the simulations. The present unbroken-sheet model cannot represent this detached state, so the predicted and simulated profiles depart in this regime. This breakup sets the high-$We$ limit of the inertial rising-sheet description and is addressed in \S\,\ref{pinch-off}.

\begin{figure}
  \centering
  \includegraphics[width=\textwidth]{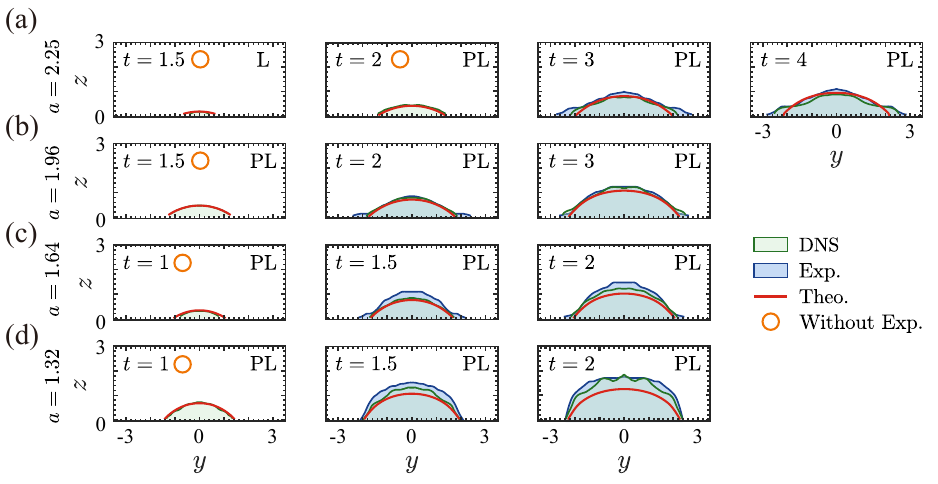}
  \caption{Side-view profiles of the central sheet at four half-spacings $a\in\{2.25,1.96,1.64,1.32\}$ (rows) and fixed $We=31$, with the post-impact time $t$ annotated in each panel. Red lines are the present theory, obtained by integrating the rim system~\eqref{eq:rim_mom_components}; blue lines are the experiments of \citet{Goswami2023} and green lines the present simulations.}
  \label{fig:appA1}
\end{figure}

We next examine the effect of the half-spacing $a\in\{2.25,1.96,1.64,1.32\}$ at fixed $We=31$. Figure~\ref{fig:appA1} compares the predicted profiles with the simulations and with the experiments of~\citet{Goswami2023}. At this lower Weber number, the central sheet is relatively shallow, and the experimental side view is often partially obscured by the drop bulk. The comparison therefore relies mainly on the simulations, while the experimental contour is shown wherever it can be extracted reliably. The predicted profiles agree closely with the simulations across all four spacings and with the experimental contours where available.

Gravity is included in the experiments and simulations but neglected in the semi-analytical model. As a result, the late-time descent of the sheet seen in the data is not reproduced. During the rising phase, however, the predicted profiles remain close to the measured and simulated contours. Together, the two comparisons span the principal control parameters, $We$ and $a$, for water at $Oh=2.9\times10^{-3}$. A further test at higher viscosity, $Oh=0.0141$, corresponding to a $40\%$ glycerol--water mixture and lying slightly outside the low-viscosity range assumed in the derivation, is reported in Appendix~\ref{app:validation}; close agreement is again obtained.

These comparisons show that the semi-analytical solution reproduces the full side-view shape over the parameter range of interest. In many applications, however, the main quantity of interest is the apex height $H(t)$ along the symmetry plane $y=0$, which is also the principal observable reported in experiments. We therefore next reduce the model to the centreline rim trajectory. On this centreline, the sheet fields simplify to closed-form expressions, and the condition for the maximum height reduces to a single algebraic relation.

\section{Maximum apex height on the centreline}
\label{sec:sheet}

\subsection{Lamella-fed stage}
\label{subsec:Lstage}

On the symmetry plane $y=0$, the formulation of \S\,\ref{sec:3d} reduces to a one-dimensional problem. The lateral motion vanishes, $\bar V=0$, the rim normal is aligned with the $z$ direction, and $\mathscr{r}=a$. The vertical velocity field is therefore inherited directly from the lamella-fed solution, $\bar W(z,t)=(z+\phi_L a)/t$, with $\partial\bar W/\partial z=1/t$.

This vertical field determines the stretching of material elements in the $z$ direction. The centreline reduction must also retain the thinning caused by lateral stretching in the $y$ direction. At the base of the sheet, this lateral stretching rate is $\partial\bar V/\partial y|_{z=0}=\phi_L/t$. In the full three-dimensional field it varies weakly with height as $(z+\phi_L a)/[t(a+z)]$. For the centreline model, we approximate it by its base value, $\partial\bar V/\partial y\equiv\phi_L/t$. This approximation keeps the solution algebraic while remaining close to the full three-dimensional rate over the range of heights relevant here. The centreline continuity equation is then
\begin{equation}
  \frac{\partial T}{\partial t} + \frac{\partial(\bar W T)}{\partial z} + \frac{\phi_L\,T}{t} = 0,
  \label{eq:sheet_cont}
\end{equation}
where the last term represents thinning by lateral stretching. Along the characteristics $\mathrm{d}z/\mathrm{d}t=\bar W$, this gives $DT/Dt=-(1+\phi_L)\,T/t$. Hence $T\,t^{1+\phi_L}$ is conserved along each characteristic, and
\begin{equation}
  T(z,t) = T_b\!\left(\frac{\phi_L\,a}{z+\phi_L\,a}\right)^{\!1+\phi_L},
  \label{eq:Tfield_I}
\end{equation}
{with $T_b=2h(a,\tau)$ is the thickness of the sheet at the base $z=0$.} The exponent $1+\phi_L$ contains a unit contribution from vertical stretching and a contribution $\phi_L$ from lateral divergence. It is the centreline analogue of the horizontal invariant $rh_at$ in the single-drop lamella solution of \citet{Gordillo2019}.

With $\bar V=0$ and the rim normal along $z$, the rim balances of \S\,\ref{sec:3d} reduce to
\begin{equation}
  \frac{\pi}{4}\frac{\mathrm{d}B^2}{\mathrm{d}t} = [\bar W(H,t)-W_r]\,T(H,t),
  \qquad
  \frac{\pi B^2}{4}\frac{\mathrm{d}W_r}{\mathrm{d}t} = [\bar W(H,t)-W_r]^2\,T(H,t) - \frac{2}{We},
  \label{eq:rim_sheet}
\end{equation}
with $W_r=\mathrm{d}H/\mathrm{d}t$. Integrating these equations gives the apex trajectory $H(t)$.

For the maximum height, the full trajectory can be reduced to a local balance at the apex. At the instant $t_m$ when the rim reaches its maximum height, $W_r=0$. We approximate this instant by a quasi-steady stall condition, in which the upward momentum flux supplied by the sheet balances the capillary retraction of the rim. This gives
\begin{equation}
  \bar W(H_m,t_m)^2\,T(H_m,t_m)\approx\frac{2}{We},
  \label{eq:stall}
\end{equation}
which we refer to as the stall condition throughout. Writing $\Gamma \equiv H_m+\phi_L a $, the apex speed is $\bar W(H_m,t_m)=\Gamma/t_m$, and the characteristic reaching $z=H_m$ at $t_m$ left the substrate at the injection time $\tau=\phi_L a\,t_m/\Gamma$, with base thickness
\begin{equation}
  T_b = \frac{18}{a^4}\left(\frac{\phi_L a\,t_m}{\Gamma}\right)^{2} h_a\!\left(\frac{3\phi_L^2 t_m^2}{\Gamma^2}\right),
  \label{eq:Tb_stall}
\end{equation}
where $h_a$ is the potential-flow thickness function of \S\,\ref{sec:single}. Substituting into the stall condition gives
\begin{equation}
  \frac{18\,\phi_L^{\,3+\phi_L}\,a^{\phi_L-1}}{\Gamma^{1+\phi_L}}\,h_a\!\left(\frac{3\phi_L^2 t_m^2}{\Gamma^2}\right) = \frac{2}{We}.
  \label{eq:algebraic_I}
\end{equation}
This relation links the maximum height $H_m=\Gamma-\phi_L a$ to the time $t_m$. The time $t_m$ is obtained from the centreline rim integration, while the remaining quantities in the stall condition are explicit. Thus the prediction is semi-analytical: the centreline fields and the stall balance are closed in algebraic form, with only the weak dependence through $h_a$ evaluated numerically. If $h_a$ were constant, $t_m$ would cancel and the height would obey the explicit scaling
\begin{equation}
  H_m \sim We^{\,1/(1+\phi_L)},
  \label{eq:Hmax_scaling}
\end{equation}
recovering $We^{1/2}$ in the inviscid limit $\phi_L\to1$. Over the present range of $a$ and $We$, however, the variation of $h_a$ is not negligible. We therefore retain the full relation~\eqref{eq:algebraic_I} and use~\eqref{eq:Hmax_scaling} only as an asymptotic guide.

\subsection{Post-lamella stage}
\label{subsec:PLstage}

Once the injection time of the apex parcel exceeds $t_\ell$, the sheet is no longer supplied by the lamella. The centreline reduction of the previous subsection still applies, but the inlet conditions are now those of the post-lamella source. On the centreline, the boundary conditions~\eqref{BC_Sheet} reduce to $\bar W(0,t)=\bar u_{PL}(t)$ and $T(0,t)=T^P(t)$. This source does not lead to a closed field of the form~\eqref{eq:Tfield_I}. Nevertheless, the apex still evolves along the vertical generator $y=0$, and the characteristic construction can be carried out on this line.

A parcel released at time $\tau$ rises ballistically with the velocity acquired at injection, where we assume velocity has the form of $\bar u_{PL}(\tau)=C\tau^{-\zeta}$ based on the full simulation results as shown in Appendix~\ref{app:inlet}. Its position is therefore $z=C\tau^{-\zeta}(t-\tau)$. Eliminating $\tau$ gives the vertical stretching rate $\partial\bar W/\partial z=\zeta/[\zeta t-(\zeta-1)\tau]$, compared with $1/t$ in the lamella-fed stage. The lateral divergence also differs from the lamella-fed case. It is now set by the source value $\partial\bar V/\partial y|_{y=0}=\bar u_{PL}(t)/a=C/(a\,t^{\zeta})$, rather than by the constant rate $\phi_L/t$. Integrating~\eqref{eq:T_characteristic_general} along the vertical generator gives
\begin{equation}
  T(z,t)=T^P(\tau)\,\frac{\tau}{\zeta t-(\zeta-1)\tau}\,\exp\!\left[-\frac{C}{a(1-\zeta)}\left(t^{1-\zeta}-\tau^{1-\zeta}\right)\right],
  \label{eq:T_axis_II}
\end{equation}
where the rational prefactor accounts for vertical stretching and the exponential factor accounts for lateral thinning. Thus, unlike the lamella-fed result~\eqref{eq:Tfield_I}, the post-lamella thickness does not reduce to a simple power law.

The maximum height is again estimated from the stall condition. The apex parcel travels from its injection time $\tau_m$ to the time of maximum height $t_m$ with the constant velocity acquired at injection. We define the flight-time ratio
\begin{equation}
  \eta \equiv \;\frac{t_m-\tau_m}{\tau_m},
  \label{eq:eta_def}
\end{equation}
so that $\tau_m=t_m/(1+\eta)$. The corresponding apex height is
\begin{equation}
  H_m=C\,\eta\,(1+\eta)^{\zeta-1}\,t_m^{\,1-\zeta}.
  \label{eq:Hm_II}
\end{equation}
{Evaluating the post-lamella field~\eqref{eq:T_axis_II} at the apex gives the thickness of the sheet arriving at the rim}
\begin{equation}
  T_r\equiv T(H_m,t_m)=\frac{T^P\!\left(t_m/(1+\eta)\right)}{1+\zeta\eta}\,\exp\!\left[-\frac{C\,t_m^{\,1-\zeta}}{a(1-\zeta)}\left(1-(1+\eta)^{-(1-\zeta)}\right)\right],
  \label{eq:Tr_II}
\end{equation}
and the incoming apex speed is $\bar W_m=C(1+\eta)^{\zeta}t_m^{-\zeta}$. Substitution into the stall condition~\eqref{eq:stall} gives
\begin{equation}
  We\;=\;\frac{2\,(1+\zeta\eta)\,t_m^{\,2\zeta}}{C^{2}\,(1+\eta)^{2\zeta}\,T^P\!\left(t_m/(1+\eta)\right)}\,\exp\!\left[\frac{C\,t_m^{\,1-\zeta}}{a(1-\zeta)}\left(1-(1+\eta)^{-(1-\zeta)}\right)\right].
  \label{eq:We_eta}
\end{equation}
Equation~\eqref{eq:We_eta} determines the flight-time ratio $\eta$ implicitly. For a given Weber number, $\eta$ is found numerically, with $t_m$ supplied by the centreline rim integration as in the lamella-fed stage; the maximum height then follows from~\eqref{eq:Hm_II}. In contrast to the lamella-fed stage, the post-lamella stage does not yield a closed power-law dependence of $H_m$ on $We$. Both the vertical-stretching prefactor and the lateral-thinning exponential vary over the present parameter range, and the Weber-number dependence enters implicitly through $\eta$. This behaviour is consistent with the absence of a single robust exponent for the apex height in the available experiments and simulations.

\subsection{Comparison with experiment and simulation}
\label{subsec:compare}

\begin{figure}
  \centering
  \includegraphics[width=\textwidth]{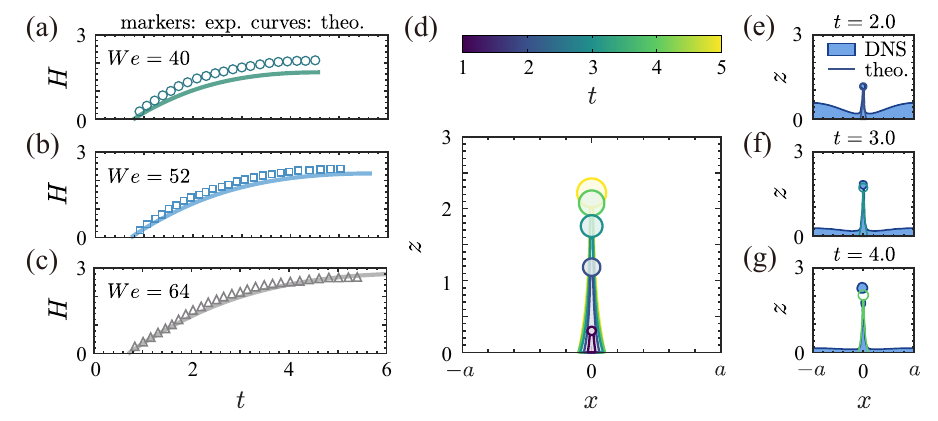}
  \caption{All results in this figure correspond to water ($Oh=2.9\times 10^{-3}$) at $a=1.80$. (\textit{a}--\textit{c})~Time evolution of the sheet apex height $H(t)$ for $We=40$, $52$ and $64$; symbols denote the experimental measurements of~\citet{Goswami2023} and solid curves the present model, obtained by integrating the rim system~\eqref{eq:rim_sheet} in time. (\textit{d})~Predicted cross-sectional profiles of the central rising sheet at $We=52$ and successive times $t=1$ to $5$, colour-coded by the horizontal colour bar. (\textit{e}--\textit{g})~Comparison between the predicted profile (curve) and direct numerical simulation (blue shaded region) at $t=2.0$, $3.0$ and $4.0$, respectively. The rim has pinched off in (\textit{g}).}
  \label{fig:sheet_evolution}
\end{figure}

Together, the lamella-fed and post-lamella reductions of \S\,\ref{subsec:Lstage}--\ref{subsec:PLstage} predict the apex height $H_m$ over the full range of impact conditions. The maximum height is only a scalar summary of the impact, however, and a more stringent test is whether the model also reproduces the time-resolved rise and the cross-sectional structure of the sheet. We carry out this test at $a=1.80$, the value used by both \citet{Goswami2023} and \citet{Zhang2026}. Figure~\ref{fig:sheet_evolution}\textit{a--c} compares the predicted $H(t)$ with the measurements of~\citet{Goswami2023} at $We=40$, $52$ and $64$; the predicted curves, obtained by integrating the rim system~\eqref{eq:rim_sheet} in time, reproduce the rise, the stall, and the approach to $H_m$ at all three Weber numbers without adjustable parameters. Figure~\ref{fig:sheet_evolution}\textit{d} shows the predicted cross-section at $We=52$ for $t=1$ to $5$, tracing the sheet as it narrows and climbs. Figure~\ref{fig:sheet_evolution}\textit{e--g} overlays this profile on the DNS field at $t=2.0$, $3.0$ and $4.0$. At $t=2.0$ and $3.0$ the agreement is excellent: the model captures not only the apex height but the full shape of the sheet, a continuous slender column tapering from the broad base to the rim. By $t=4.0$, however, the apex has pinched off in the simulation, shedding a detached droplet above a base column that has stalled below it. The unbroken-sheet description, which assumes a continuous column from base to apex, cannot represent this configuration. The same detachment is observed in both experiment and simulation as $We$ increases, and marks the limit of the present model: the central sheet eventually breaks up under capillary action, the regime we turn to next.

\section{Capillary pinch-off and the limit of the model}
\label{pinch-off}

\subsection{Capillary breakup and the apex cut-off}
\label{subsec:RP}

\begin{figure}
    \centering
    \includegraphics[width=\linewidth]{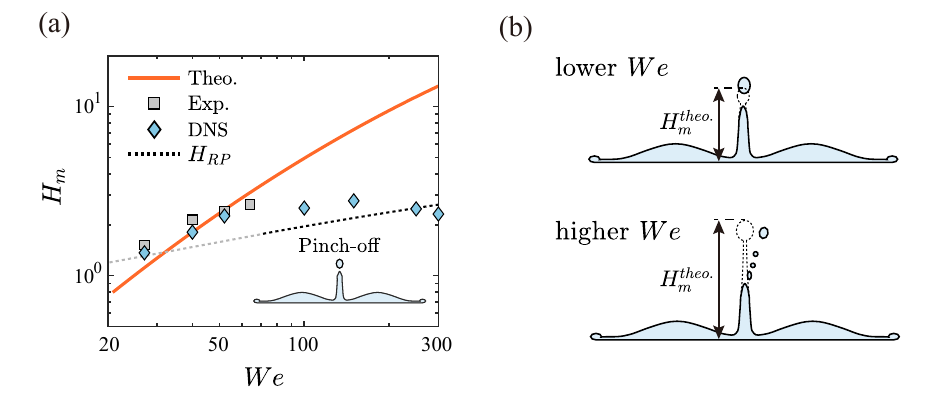}
    \caption{Maximum central-sheet height $H_m$ versus Weber number $We$ at $a=1.80$ and $Oh=2.9\times10^{-3}$. (\textit{a})~The solid line is the inertial prediction~\eqref{eq:We_eta} and the dotted line the breakup-limited height $H_{RP}$~\eqref{eq:H_RP}. Grey squares are the experiments of \citet{Goswami2023} and blue diamonds the present DNS. (\textit{b})~Schematic of the two regimes; the dashed extension marks the inertial height $H_m^{\mathrm{theo.}}$ the sheet would reach without breakup. At lower $We$ (top) the sheet reaches $H_m^{\mathrm{theo.}}$ before pinch-off; at higher $We$ (bottom) it pinches during the rise and stalls at $H_{RP}$.}
    \label{pinchoff}
\end{figure}
To capture this breakup, we extend the model with a capillary cut-off built on the thinning sheet itself. As the sheet rises it thins, and at high $We$ the post-lamella stage draws it out into a slender jet that is unstable to capillary perturbations; in the lamella-fed stage the sheet is still thick, fed, and rim-bounded, and no such instability develops. The analysis therefore proceeds along the post-lamella characteristics, on which the thickness is already known in closed form from~\eqref{eq:T_axis_II}, so a linear Rayleigh--Plateau treatment of the jet introduces no further parameters. The growth of these perturbations sets a breakup time, beyond which the detached fragments carry vertical momentum out of the sheet and the apex can rise no higher; the resulting height $H_{RP}$ caps the inertial prediction at high $We$.

{
We follow the parcel injected on the axis at $\tau$ along its characteristic; the local jet radius is $r_{\mathrm{jet}}(\tau,t)=T(\tau,t)/2$, decreasing monotonically with $t$. Because the thinning rate $T^{-1}\partial T/\partial t = O(1)$ stays well below the capillary growth rate $\omega_{RP}=O(10)$ wherever breakup is binding, the most-amplifying mode at each instant is set by the local radius, and a quasi-steady dispersion relation applies pointwise along the characteristic. Taking the viscous Rayleigh relation~\citep{Rayleigh1878,Chandrasekhar1961,Eggers2008} at the inviscid most-unstable wavenumber $kr_{\mathrm{jet}}=0.697$,
\begin{equation}
  \omega_{RP}(\tau,t) \;=\; \sqrt{\mathcal{D}^{2}+\mathcal{G}^{2}} -\mathcal{D}, \qquad \mathcal{D} \;=\; \frac{0.729}{Re\,r_{\mathrm{jet}}^{2}}, \qquad \mathcal{G} \;=\; \frac{0.343}{\sqrt{We\,r_{\mathrm{jet}}^{3}}},
  \label{eq:sigma_RP}
\end{equation}
with $\mathcal{D}$ the viscous damping and $\mathcal{G}$ the inviscid Rayleigh growth; $0.343$ is the classical~\citet{Rayleigh1878} maximum growth rate and $0.729=1.5\times(0.697)^{2}$ its viscous counterpart. Following~\citet{Weber1931}, breakup occurs when a perturbation grows from an initial amplitude $\varepsilon_0$ to one comparable with the jet radius,
\begin{equation}
  \int_{\tau}^{t_{\mathrm{brk}}} \omega_{RP}(\tau,\tau')\,\mathrm{d}\tau' \;=\; \ln(r_{\mathrm{jet}}/\varepsilon_0),
  \label{eq:Weber_threshold}
\end{equation}}
with $\ln(r_{\mathrm{jet}}/\varepsilon_0)\approx12$, the standard value for capillary jet breakup and within the range $8$--$15$ reported in the literature. Equation~\eqref{eq:Weber_threshold} sets the breakup time $t_{\mathrm{brk}}$, and the parcel trajectory then fixes the breakup-limited apex height,
\begin{equation}
  H_{RP} \;=\; \bar u_{PL}(\tau)\,(t_{\mathrm{brk}} - \tau), \qquad \bar u_{PL}(\tau) \;=\; C\,\tau^{-\zeta}.
  \label{eq:H_RP}
\end{equation}
Every quantity on the right of~\eqref{eq:sigma_RP}--\eqref{eq:H_RP} is either an inlet condition, an impact parameter ($We$, $Re$), or the literature threshold; none is adjusted to the comparisons below.

\subsection{Comparison and range of validity}
\label{subsec:Hmax_compare}

\begin{figure}
  \centering
  \includegraphics[width=\textwidth]{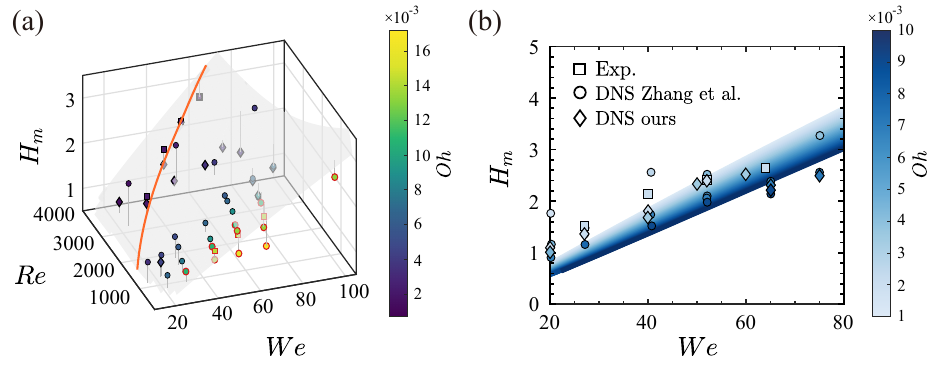}
  \caption{Maximum central-sheet height $H_m$ over the $(We,Re)$ plane at $a=1.80$. (\textit{a})~The grey surface is the inertial prediction~\eqref{eq:We_eta}; the orange curve is its intersection with the iso-$Oh$ contour at $Oh=2.9\times10^{-3}$ (water), identical to the model curve in figure~\ref{pinchoff}\textit{a}. Diamonds are the present simulation and squares the experiments of \citet{Goswami2023,Goswami2026}; filled circles are the simulation results of \citet{Zhang2026}, coloured by $Oh$ (colour bar).Red-edged circles mark $Oh\geq0.01$, outside the low-viscosity range of the derivation. (\textit{b})~The same comparison restricted to $10^{-3}\lesssim Oh\lesssim10^{-2}$, where the model assumptions hold; the blue band is the prediction across this range, coloured by $Oh$, and symbols follow~(\textit{a}).}
  \label{fig:Hmax_eta}
\end{figure}
We first follow a single liquid (water, $Oh=2.9\times10^{-3}$) to high $We$, beyond the experimentally accessible range. Figure~\ref{pinchoff}\textit{a} compares the inertial prediction~\eqref{eq:We_eta} (solid), its lamella-fed asymptote~\eqref{eq:Hmax_scaling} (triangle), the breakup-limited estimate $H_{RP}$~\eqref{eq:H_RP} (dotted), the experiments of~\citet{Goswami2023} (squares), and the present DNS (diamonds). The data interpolate smoothly between the two predictions. At lower $We$ the apex jet, if it forms at all, breaks only after the sheet has essentially reached its maximum; little vertical momentum is lost, and $H_m$ follows the inertial prediction~\eqref{eq:We_eta}, which we denote $H_m^{\mathrm{theo.}}$ as the height the sheet would attain in the absence of breakup. As $We$ rises the jet is drawn out earlier and thinner, breakup sets in during the rise, and the successive shed drops drain momentum from the apex, so that $H_m$ peels away from $H_m^{\mathrm{theo.}}$ toward $H_{RP}$. Figure~\ref{pinchoff}\textit{b} sketches the two regimes observed in our simulations: at lower $We$ breakup is late and releases at most one drop, so $H_m\approx H_m^{\mathrm{theo.}}$; at higher $We$ the jet pinches repeatedly during the rise, leaving $H_m$ well below $H_m^{\mathrm{theo.}}$.

Within the lower-$We$ range for which the model is built, pinch-off thus has only a modest effect on $H_m$. To confirm that the inertial prediction holds across this range as $Oh$ and $Re$ vary, we compare it against experiments and simulations over the full $(We,Re)$ plane. In figure~\ref{fig:Hmax_eta}\textit{a} the grey surface is the inertial prediction, computed on a grid of $1600$ points spanning the plane; the orange curve is its intersection with the $Oh=2.9\times10^{-3}$ contour, identical to the model curve of figure~\ref{pinchoff}\textit{a}. Diamonds (present simulation) and squares (experiments of \citealt{Goswami2023,Goswami2026}) follow the colour convention of the ensemble, and red-edged circles flag $Oh \geq 0.01$, outside the low-viscosity range of the derivation. Restricting the comparison to $10^{-3}\lesssim Oh\lesssim10^{-2}$, where the model assumptions hold, the prediction agrees with both experiment and simulation to within about $\pm20\%$ (figure~\ref{fig:Hmax_eta}\textit{b}). 

This agreement, however, holds only within the parameter range over which the derivation is valid. At high $We$ two distinct effects appear. Capillary breakup of the rising sheet is already accounted for through $H_{RP}$; the single-drop lamella, however, may itself splash before reaching the collision line~\citep{Riboux2014}, corrupting the inlet upstream of the rim collision in a way the present description does not capture. At low $Oh$, the inertial derivation rests on a thin viscous correction to a free-slip lamella, which loses accuracy as $Oh$ approaches $10^{-2}$ and the boundary layer occupies an appreciable fraction of the lamella thickness. Between these limits the theory gives a self-consistent quantitative account of the central-sheet dynamics, from inertial feeding and two-directional thinning to capillary-limited expansion and the Rayleigh--Plateau closure, with no parameter adjusted beyond the once-calibrated post-lamella inlet.

\section{Conclusions}
\label{sec:conclusions}

We have developed a semi-analytical framework for the central rising sheet produced by the simultaneous impact of two drops on a solid substrate. The model couples the thin-film flow in the sheet to the bounding rim. It gives the velocity and thickness fields in the sheet and, through the centreline rim dynamics, the apex trajectory \(H(t)\).

The framework describes the sheet through two feeding stages. During the lamella-fed stage, \(t_c\leq t\leq t_\ell\), the inlet flow is inherited from the single-drop lamella solution. The simple radial form of this inlet allows the ballistic characteristic construction to be carried out analytically. On the centreline, the vertical stretching and lateral divergence combine to give the conserved quantity \(Tt^{1+\phi_L}\), and the maximum height is reduced to an algebraic balance between inertial feeding and capillary retraction. The exponent \(1+\phi_L\) reflects the combined thinning in the vertical and lateral directions. After \(t_\ell\), the lamella no longer reaches the collision line. The sheet is then supplied by the remaining bulk motion of the two drops, and the inlet no longer has a closed analytical form. We therefore prescribe the post-lamella inlet from simulations and continue the same characteristic construction based on the empirical boundary conditions. With this inlet, the model predicts the sheet trajectory and shape in good agreement with experiments and simulations over the parameter range considered.

The model also identifies the high-\(We\) limit of the unbroken-sheet description. At sufficiently large Weber number, the apex thins into a slender jet and pinches off before the inertial sheet would reach its predicted maximum height. We account for this cutoff using a linear Rayleigh--Plateau analysis, with the time-dependent jet diameter and deceleration supplied by the sheet model. The resulting upper bound \(H_{RP}\) agrees well with both experiments and simulations.

The main limitation of the framework lies in the post-lamella inlet. In the lamella-fed stage, the inlet is determined by the single-drop lamella solution, and the injection point, injection time, velocity field, and thickness field can all be obtained explicitly. In the post-lamella stage, by contrast, the inlet is set by the residual bulk motion after the lamella front has passed the collision line. This flow does not reduce to a simple radial inlet law, and the inverse characteristic map cannot be written in closed form. The fitted post-lamella inlet used here captures the dynamics over the full range of conditions considered in this study, but it should not be regarded as a universal relation. A substantially different parameter range would require a new calibration.

This limitation concerns the inlet to the sheet rather than the subsequent sheet dynamics. Once the inlet velocity and thickness are specified, the rest of the model is deterministic: fluid parcels follow ballistic characteristics, the thickness evolves by mass conservation, the rim motion follows from mass and momentum balances, and the breakup cutoff is determined by the Rayleigh--Plateau criterion. The framework therefore separates the part of the problem that can be treated analytically from the single empirical input required in the post-lamella stage.

Although we have considered the simultaneous impact of two identical drops, the same construction can be extended to more general configurations. Natural next steps include non-simultaneous impacts and unequal drop sizes, which are closer to the conditions encountered in spray cooling, coating, and inkjet printing.

\noindent\textbf{Supplementary movies.}\ Movies comparing the semi-analytical predictions with the numerical simulations over a range of parameters are provided as supplementary material.

\noindent\textbf{Acknowledgements.}\ S.S.H.\ thanks DeepSeek for assistance with spell-checking and grammar refinement of the manuscript. N.H. thanks Tachin Ruangkriengsin for the insightful discussion.

\noindent\textbf{Funding.}\ L.W.F. acknowledges the support from the National Natural Science Foundation of China (Grant No.\ 52276088).

\noindent\textbf{Declaration of interests.}\ The authors report no conflict of interest.

\appendix
\section{Lamella dynamics from a single impacting drop}
\label{sec:single_appendix}

Within the lamella, the depth-averaged radial velocity $\bar u(r,t)=\frac{1}{h}\int_0^hu(r,z,t)\mathrm{d}z$ and film thickness $h(r,t)$ satisfy the thin-film conservation laws
\begin{equation}
  \frac{\partial(rh)}{\partial t}+\bar u\frac{\partial(rh)}{\partial r}=-rh\frac{\partial \bar u}{\partial r},
  \label{eq:lam_cont}
\end{equation}
\begin{equation}
  \frac{\partial \bar u}{\partial t}+\bar u\frac{\partial \bar u}{\partial r}=-\frac{\lambda\, \bar u}{h\sqrt{Re\,t}},
  \label{eq:lam_mom}
\end{equation}
with $\lambda=1$ \citep{Gordillo2019}.  The right-hand side of \eqref{eq:lam_mom} represents wall friction from the substrate boundary layer, whose thickness grows as $\delta=(t/Re)^{1/2}$ in the stagnation-point regime \citep{Roisman2009,Eggers2010}.  At the inner boundary $r=\sqrt{3t}$ the lamella is fed by the collapsing drop with the boundary conditions
\begin{equation}
\bar  u\!\left(\sqrt{3t},t\right)=\sqrt{3/t},\qquad h\!\left(\sqrt{3t},t\right)=h_{a}(t),
  \label{eq:lam_BC}
\end{equation}
where $h_{a}(t)$ is a known function, independent of $We$ and $Re$, tabulated as a ninth-order polynomial fit to the potential-flow solution~\citep{Gordillo2019}.

Since $Re \gg 1$, we expand the velocity as $\bar u = \bar u_0 + Re^{-1/2}\bar u_1 + O(Re^{-1})$. At leading order, the friction term is negligible, so $\bar u_0$ remains constant along each characteristic. Let $\tau$ denote the time at which a fluid parcel enters the lamella. At that instant, the inner edge of the lamella is located at $r=\sqrt{3\tau}$, and the boundary condition gives the parcel velocity $\bar u_0=\sqrt{3/\tau}$. Because $\bar u_0$ remains constant along the parcel trajectory, the parcel position at a later time $t$ satisfies $r=\sqrt{3\tau}+\sqrt{3/\tau}(t-\tau)=\sqrt{3/\tau}\,t$. Solving for $\tau$ gives $\tau=3t^2/r^2$, and therefore
\begin{equation}
  \bar u_0(r,t)=\sqrt{3/\tau}=r/t
  \label{eq:u0}
\end{equation}
This self-similar velocity field is a purely kinematic consequence of the boundary condition \eqref{eq:lam_BC}, independent of $We$, $Re$ or the details of $h_{a}$.  The leading-order continuity equation yields the invariant $D(rh_{0}t)/Dt=0$, giving the thickness profile based on $r h_0 t = \sqrt{3\tau}\,h_a(\tau)\,\tau$ as follows:
\begin{equation}
  h_{0}(r,t)=\frac{9t^{2}}{r^{4}}\,h_{a}\!\!\left(\frac{3t^{2}}{r^{2}}\right).
  \label{eq:h0}
\end{equation}
The first-order viscous correction, derived in full by \citet{Gordillo2019}, is
\begin{equation}
  \bar u_{1}(r,t)=-\frac{1}{t\,h_{a}(x)}\left[\frac{\sqrt{3}\chi\, \tau}{2}+\frac{2\sqrt{3}\lambda}{7\tau^{5/2}}\left(t^{7/2}-\tau^{7/2}\right)\right],\qquad \tau=\frac{3t^{2}}{r^{2}},
  \label{eq:u1}
\end{equation}
with $\chi=0.6$.  It is convenient to define a viscous attenuation factor $\phi$ by writing
\begin{equation}
  \bar u(r,t)=\phi(r,t)\,\frac{r}{t},\qquad \phi\equiv 1+\frac{\bar u_{1}}{\bar u_{0}}\,Re^{-1/2}+O(Re^{-1})\le 1,
  \label{eq:phi_def}
\end{equation}
so that $\phi<1$ measures the fraction of the inviscid leading-order velocity that survives viscous retardation. Because $\phi$ depends on position and time only through the combination $x=3t^{2}/r^{2}$, it is constant along each leading-order characteristic and varies slowly across the lamella at any instant. Over the parameter range considered it depends mainly on $Oh$ and only weakly on $We$, changing by less than about $10\%$ across the experimental $We$ range at fixed $Oh$. We therefore treat $\phi$ as a constant, denoted $\phi_L$, read off from the numerical solution for each $(We,Oh)$ and used as a fixed input to the lamella-fed stage in §\ref{Dynamics and stages description}; the post-lamella stage instead takes its inlet directly from simulation, as described there.

The rim at $r=s(t)$ advances at a velocity of $u_r\equiv ds/dt$ and is governed by the mass and momentum balance equations
\begin{equation}
  \alpha\frac{\pi}{4}\frac{\mathrm{d}b^{2}}{\mathrm{d}t}
  =\bigl(\bar u(s,t)-u_r\bigr)h(s,t),\qquad 
  \alpha\frac{\pi b^{2}}{4}\frac{\mathrm{d}u_r}{\mathrm{d}t}
  =\bigl(\bar u(s,t)-u_r\bigr)^{2}h(s,t)-\frac{1+\beta}{We},
  \label{eq:rim_single}
\end{equation}
where $\alpha$ and $\beta$ depend on the substrate wettability~\citep{Gordillo2019}. For a hydrophilic substrate $\alpha=1/2$ and $\beta=-\cos\theta$, with $\theta$ being the advancing contact angle. Eqns.~\eqref{eq:lam_cont}--\eqref{eq:rim_single}, together with the initial conditions at the ejection time $t_{e}=1.05\,We^{-2/3}$ specified by~\citet{Riboux2014,Riboux2015}, form the complete single-drop problem. The rim ODE is integrated numerically; the lamella fields upstream of the rim are evaluated from the validated expressions~\eqref{eq:u0}--\eqref{eq:u1}. The resulting numerical solutions provide the time-dependent rim position $s(t)$ together with the lamella state at arbitrary $(r,t)$, both of which serve as input for the pair-impact analysis that follows.

\begin{figure}
  \centering
  \includegraphics[width=\textwidth]{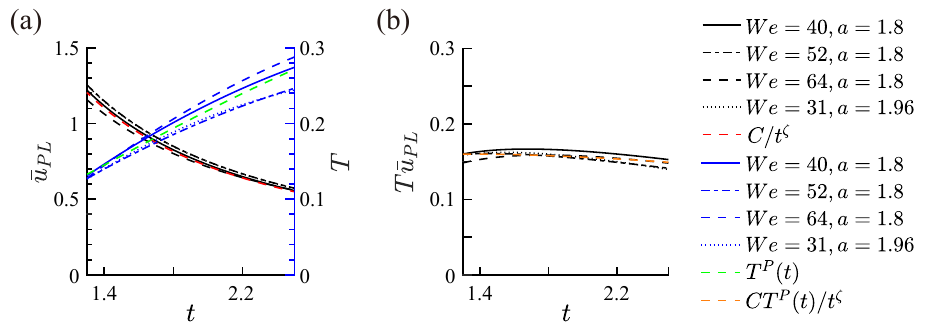}
  \caption{Post-lamella inlet for four representative cases ($We=40,52,64$ at $a=1.8$, and $We=31$ at $a=1.96$). (\textit{a})~Injection velocity $\bar u_{PL}$ (black, left axis) and thickness $T$ (blue, right axis); the red and green curves are the fitted forms $C/t^{\zeta}$ and $T^P(t)$. (\textit{b})~Injection flux $T\bar u_{PL}$. Line styles denote the four cases.}
  \label{fig:inlet}
\end{figure}

\section{Boundary conditions of the post-lamella stage}
\label{app:inlet}

Once the lamella no longer feeds the sheet, the inflow is set by the bulk converging beneath the colliding rims, a flow that has no closed-form description and presents the same obstruction that has long kept the single-drop spreading problem open. We therefore fix the post-lamella inlet from the simulations, following the same route by which \citet{Gordillo2019} close their lamella solution with a thickness function $h_a$ read from the numerical potential-flow solution. We extract the inlet thickness and injection velocity from the simulations and represent them by simple functions of time.

Both quantities are measured along the collision line. The rising sheet stands on the converging bulk beneath the rims, and the two meet at a sharp junction where the liquid narrows from the wide bulk below into the thin sheet, which we take as the base of the sheet. The inlet thickness $T$ is the width of the sheet at this junction and the injection velocity $\bar u_{PL}$ the mean vertical velocity across it, each averaged with the local liquid fraction. The same construction is applied to every case and every instant, so that $T$ and $\bar u_{PL}$ are obtained on a common footing and need no case-by-case adjustment.

Figure~\ref{fig:inlet} shows the extracted $\bar u_{PL}$ and $T$. Both vary only weakly between cases and show no systematic trend with $We$ or $a$. {The velocity roughly follows the decaying power law $\bar u_{PL}=C/t^{\zeta}$ with $\zeta=1.2$ taken to be the same for every case,} the prefactor $C$ being set by continuity with the lamella-fed velocity at $t_{\ell}$ rather than by fitting. The thickness follows a cubic in time, again common to all cases, denoted $T^P(t)$ in~\eqref{BC_Sheet} and given by
\begin{equation}
  T^P(t)=-0.00396\,t^{3}+0.01298\,t^{2}+0.11063\,t-0.02427.
  \label{eq:Pt}
\end{equation}

Although $\bar u_{PL}$ and $T$ each vary somewhat between cases, their product $T\bar u_{PL}$ does not, because a thicker inlet is accompanied by a slower injection and the two variations cancel. It is this product, and not $\bar u_{PL}$ or $T$ separately, that fixes the momentum delivered to the rim, so the downstream solution depends on the inlet only through $T\bar u_{PL}$ and is largely insensitive to the case-to-case differences in the two factors. This robustness comes with a corresponding limitation. The inlet is obtained by fitting, as indeed is the thickness function $h_a$ of the single-drop solution, but the two differ in range: $h_a$ is independent of $We$ and $Re$ and so holds at any impact condition, whereas a fitted inlet holds over a finite window of conditions and is not universal. The present fit covers the full range examined in this work, and a sufficiently different regime would require its own.

\section{Direct numerical simulations}
\label{sec:dns_appendix}

The analytical model developed in the main text relies on a closed-form description of the lamella feed, whose validity over the impact-parameter range of interest is best assessed against three-dimensional simulations of the full Navier--Stokes equations. We carry out such simulations in \textsc{Basilisk}~\citep{Popinet2009,Popinet2018}, the same open-source platform used by~\citet{Zhang2026} for the simultaneous pair-drop impact problem, and adopt their numerical setup essentially without modification.

{
The two phases are governed by the incompressible Navier--Stokes equations with surface tension, in the dimensionless form introduced in \S\,\ref{sec:single} (lengths scaled by $R^*$, velocities by $V^*$, time by $R^*/V^*$, and pressure by $\rho^*V^{*2}$),
\begin{equation}
  \nabla\cdot\mathbf u=0,
  \label{eq:continuity_dns}
\end{equation}
\begin{equation}
  \rho\left(\frac{\partial\mathbf u}{\partial t}+\mathbf u\cdot\nabla\mathbf u\right)=-\nabla p+\frac{1}{Re}\,\nabla\cdot\!\big[\mu(\nabla\mathbf u+\nabla\mathbf u^{\mathsf T})\big]+\frac{1}{We}\,\kappa\,\delta_s\,\mathbf n-\frac{1}{Fr^{2}}\,\rho\,\mathbf e_z,
  \label{eq:momentum_dns}
\end{equation}
where $\mathbf u$ is the full velocity field, $p$ the pressure, $\kappa$ and $\mathbf n$ the curvature and unit normal of the interface, and $\delta_s$ the interfacial Dirac distribution; $We$, $Re$ and $Fr$ are the groups defined in \S\,\ref{sec:single} and \S\,\ref{sec:3d}, and $\mathbf e_z$ is the upward unit vector. The dimensionless density and viscosity follow the volume fraction $f$ as $\rho=f+(1-f)/\rho_r$ and $\mu=f+(1-f)/\mu_r$, with $f=1$ in the liquid, $f=0$ in the gas, and ratios $\rho_r=\rho^*_l/\rho^*_g=813$, $\mu_r=\mu^*_l/\mu^*_g=56$ for water in air. Gravity is retained in the simulations through the body-force term; as shown in \S\,\ref{sec:3d} it is dynamically negligible over the range considered ($We^{1/2}Fr^{-2}\lesssim0.1$), so it does not materially affect the sheet dynamics.
}

The equations are solved with an octree-based adaptive mesh refinement (AMR) method and a momentum-conserving geometric volume-of-fluid (VOF) formulation; the Bell--Colella--Glaz scheme is used for advection, viscous terms are treated implicitly, and surface tension is applied through the continuum-surface-force model~\citep{Brackbill1992,Popinet2009}. We use a cubic domain of side $16$, with two identical drops of radius $1$ released a small distance above the substrate and given the impact velocity $U_0$ directed downward, at a centre-to-centre horizontal separation $2a$, where $a$ is the impact parameter of \S\,\ref{sec:3d}. The substrate at $z=0$ is a no-slip wall with a static contact angle $\theta_s=90^{\circ}$, imposed through a Neumann condition on the volume fraction, while the lateral and upper boundaries use outflow conditions~\citep{Sanjay2023a,Sanjay2023b}.

The maximum refinement level is $L_{\max}=12$, giving a minimum cell size $\Delta_{\min}\approx4\times10^{-3}$, about $250$ cells across the drop radius; refinement is driven by wavelet thresholds $f_{\mathrm{err}}=5\times10^{-5}$ on the volume fraction and $u_{\mathrm{err}}=5\times10^{-3}$ on the velocity, which keep the central sheet and rim on the finest grid throughout the rising phase. We verified that the apex height $H(t)$ is grid-converged at this resolution, consistent with the mesh-convergence tests of~\citet{Zhang2026} at the same level. The setup reproduces the experiments of~\citet{Goswami2023} across the range examined here; the present simulations at $a=1.80$ shown in figure~\ref{fig:Hmax_eta}(\textit{a}) likewise recover the same experimental dataset.

\section{Additional validation cases}
\label{app:validation}
\begin{figure}
  \centering
  \includegraphics[width=\textwidth]{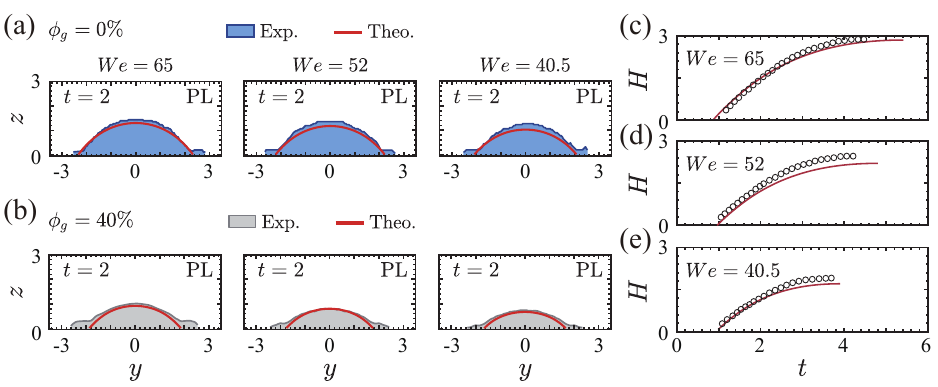}
  \caption{Validation against the glycerol--water dataset of \citet{Goswami2026} ($Oh=0.0141$). (\textit{a},\textit{b})~Side-view profiles at $t=2$ for water ($\phi_g=0\%$) and the $40\%$ glycerol--water mixture ($\phi_g=40\%$) at $We=65,52,40.5$ (columns). (\textit{c}--\textit{e})~Apex height $H(t)$ for the mixture at the same Weber numbers. Filled contours and open circles are the experiments of \citet{Goswami2026} (blue for water, grey for the mixture); red lines are the present theory.}
  \label{fig:appA2}
\end{figure}
The main-text comparisons validate the model against water across Weber number and drop spacing. To probe the remaining parameter direction, viscosity, we test it here against the $40\%$ glycerol--water dataset of \citet{Goswami2026}, for which $Oh=0.0141$. This value lies marginally above the low-viscosity range $Oh\lesssim10^{-2}$ over which our governing equations are derived, so the comparison probes the model just beyond its formal regime of validity. The remaining mixtures in that dataset, with glycerol fractions up to $80\%$, lie further outside the range of the single-drop spreading solution of \citet{Gordillo2019} on which our framework rests; \citet{Goswami2026} extend the spreading description to those cases through adjusted coefficients, but we do not pursue that extension here.

Figure~\ref{fig:appA2}\textit{a},\textit{b} compares the side-view profiles at $t=2$ for water and the glycerol--water mixture at $We=65$, $52$ and $40.5$. The added viscosity visibly reshapes the sheet: at a given Weber number the apex is markedly lower than for water, reflecting the enhanced viscous dissipation during the rise. The predicted contours nonetheless track the measured ones in both rows, so the model captures this viscous reshaping without any retuning of its parameters. Figure~\ref{fig:appA2}\textit{c}--\textit{e} follows the apex height $H(t)$ for the mixture across the full rise. The model reproduces the rate of rise at all three Weber numbers and the maximum height at $We=65$; at the two lower Weber numbers it slightly underpredicts the late-time apex, as expected when $Oh$ exceeds the range over which the lamella description is strictly valid.

Taken together, these comparisons show that the model continues to capture the central-sheet evolution for an Ohnesorge number modestly above its formal bound, the discrepancy remaining small and confined to the late stage of the rise. This supports its use across the full $Oh$ range of the main text.

\bibliography{ref}

@article{Goswami2026,
  title = {On the role of liquid viscosity during droplet-pair impacts on solid surfaces},
  volume = {1033},
  ISSN = {1469-7645},
  url = {http://dx.doi.org/10.1017/jfm.2026.11437},
  DOI = {10.1017/jfm.2026.11437},
  journal = {Journal of Fluid Mechanics},
  publisher = {Cambridge University Press (CUP)},
  author = {Goswami,  Anjan and Hardalupas,  Yannis},
  year = {2026},
  month = Apr 
}

@article{Wagner1932,
  title = {\"{U}ber Stoß‐ und Gleitvorg\"{a}nge an der Oberfl\"{a}che von Fl\"{u}ssigkeiten},
  volume = {12},
  ISSN = {1521-4001},
  url = {http://dx.doi.org/10.1002/zamm.19320120402},
  DOI = {10.1002/zamm.19320120402},
  number = {4},
  journal = {ZAMM - Journal of Applied Mathematics and Mechanics / Zeitschrift f\"{u}r Angewandte Mathematik und Mechanik},
  publisher = {Wiley},
  author = {Wagner,  Herbrt},
  year = {1932},
  month = Jan,
  pages = {193–215}
}

@article{Sanjay2023a,
  title = {When does an impacting drop stop bouncing?},
  volume = {958},
  ISSN = {1469-7645},
  url = {http://dx.doi.org/10.1017/jfm.2023.55},
  DOI = {10.1017/jfm.2023.55},
  journal = {Journal of Fluid Mechanics},
  publisher = {Cambridge University Press (CUP)},
  author = {Sanjay,  Vatsal and Chantelot,  Pierre and Lohse,  Detlef},
  year = {2023},
  month = Mar 
}

@article{Sanjay2023b,
  title = {Drop impact on viscous liquid films},
  volume = {958},
  ISSN = {1469-7645},
  url = {http://dx.doi.org/10.1017/jfm.2023.13},
  DOI = {10.1017/jfm.2023.13},
  journal = {Journal of Fluid Mechanics},
  publisher = {Cambridge University Press (CUP)},
  author = {Sanjay,  Vatsal and Lakshman,  Srinath and Chantelot,  Pierre and Snoeijer,  Jacco H. and Lohse,  Detlef},
  year = {2023},
  month = Mar 
}

@article{Popinet2018,
  title = {Numerical Models of Surface Tension},
  volume = {50},
  ISSN = {1545-4479},
  url = {http://dx.doi.org/10.1146/annurev-fluid-122316-045034},
  DOI = {10.1146/annurev-fluid-122316-045034},
  number = {1},
  journal = {Annual Review of Fluid Mechanics},
  publisher = {Annual Reviews},
  author = {Popinet,  Stéphane},
  year = {2018},
  month = Jan,
  pages = {49–75}
}

@article{Popinet2009,
  title = {An accurate adaptive solver for surface-tension-driven interfacial flows},
  volume = {228},
  ISSN = {0021-9991},
  url = {http://dx.doi.org/10.1016/j.jcp.2009.04.042},
  DOI = {10.1016/j.jcp.2009.04.042},
  number = {16},
  journal = {Journal of Computational Physics},
  publisher = {Elsevier BV},
  author = {Popinet,  Stéphane},
  year = {2009},
  month = Sept,
  pages = {5838–5866}
}

@article{Brackbill1992,
  title = {A continuum method for modeling surface tension},
  volume = {100},
  ISSN = {0021-9991},
  url = {http://dx.doi.org/10.1016/0021-9991(92)90240-Y},
  DOI = {10.1016/0021-9991(92)90240-y},
  number = {2},
  journal = {Journal of Computational Physics},
  publisher = {Elsevier BV},
  author = {Brackbill,  J.U and Kothe,  D.B and Zemach,  C},
  year = {1992},
  month = June,
  pages = {335–354}
}

@book{Chandrasekhar1961,
  author    = {Chandrasekhar, S.},
  title     = {Hydrodynamic and Hydromagnetic Stability},
  publisher = {Oxford University Press},
  year      = {1961}
}

@article{Eggers2008,
  title = {Physics of liquid jets},
  volume = {71},
  ISSN = {1361-6633},
  url = {http://dx.doi.org/10.1088/0034-4885/71/3/036601},
  DOI = {10.1088/0034-4885/71/3/036601},
  number = {3},
  journal = {Reports on Progress in Physics},
  publisher = {IOP Publishing},
  author = {Eggers,  Jens and Villermaux,  Emmanuel},
  year = {2008},
  month = Feb,
  pages = {036601}
}

@article{Weber1931,
  title = {Zum Zerfall eines Fl\"{u}ssigkeitsstrahles},
  volume = {11},
  ISSN = {1521-4001},
  url = {http://dx.doi.org/10.1002/zamm.19310110207},
  DOI = {10.1002/zamm.19310110207},
  number = {2},
  journal = {ZAMM - Journal of Applied Mathematics and Mechanics / Zeitschrift f\"{u}r Angewandte Mathematik und Mechanik},
  publisher = {Wiley},
  author = {Weber,  Constantin},
  year = {1931},
  month = Jan,
  pages = {136–154}
}

@article{Rayleigh1878,
  title = {On The Instability Of Jets},
  volume = {s1-10},
  ISSN = {0024-6115},
  url = {http://dx.doi.org/10.1112/plms/s1-10.1.4},
  DOI = {10.1112/plms/s1-10.1.4},
  number = {1},
  journal = {Proceedings of the London Mathematical Society},
  publisher = {Wiley},
  author = {Rayleigh,  Lord},
  year = {1878},
  month = Nov,
  pages = {4–13}
}

@article{Bremond2006,
  title = {Atomization by jet impact},
  volume = {549},
  ISSN = {1469-7645},
  url = {http://dx.doi.org/10.1017/S0022112005007962},
  DOI = {10.1017/s0022112005007962},
  journal = {Journal of Fluid Mechanics},
  publisher = {Cambridge University Press (CUP)},
  author = {Bremond,  N. and Villermaux,  E.},
  year = {2006},
  month = Feb,
  pages = {273–306}
}

@article{HassonPeck1964,
  title = {Thickness distribution in a sheet formed by impinging jets},
  volume = {10},
  ISSN = {1547-5905},
  url = {http://dx.doi.org/10.1002/aic.690100533},
  DOI = {10.1002/aic.690100533},
  number = {5},
  journal = {AIChE Journal},
  publisher = {Wiley},
  author = {Hasson,  David and Peck,  Ralph E.},
  year = {1964},
  month = Sept,
  pages = {752–754}
}

@inproceedings{Barnes1999,
  author    = {Barnes, H. A. and Hardalupas, Y. and Taylor, A. M. K. P. and Wilkins, J. H.},
  title     = {An investigation of the interaction between two adjacent impinging droplets},
  booktitle = {Proceedings of the 15th International Conference on Liquid Atomisation and Spray Systems (ILASS)},
  year      = {1999},
  editor    = {Lavergne, G.},
  pages     = {1--7},
  address   = {Toulouse, France},
  publisher = {ONERA}
}

@article{Roisman2002,
  title = {Multiple Drop Impact onto a Dry Solid Substrate},
  volume = {256},
  ISSN = {0021-9797},
  url = {http://dx.doi.org/10.1006/jcis.2002.8683},
  DOI = {10.1006/jcis.2002.8683},
  number = {2},
  journal = {Journal of Colloid and Interface Science},
  publisher = {Elsevier BV},
  author = {Roisman,  I.V. and Prunet-Foch,  B. and Tropea,  C. and Vignes-Adler,  M.},
  year = {2002},
  month = dec,
  pages = {396–410}
}

@article{Ersoy2020,
  title = {Central uprising sheet in simultaneous and near-simultaneous impact of two high kinetic energy droplets onto dry surface and thin liquid film},
  volume = {32},
  ISSN = {1089-7666},
  url = {http://dx.doi.org/10.1063/1.5135029},
  DOI = {10.1063/1.5135029},
  number = {1},
  journal = {Physics of Fluids},
  publisher = {AIP Publishing},
  author = {Ersoy,  Nuri Erdem and Eslamian,  Morteza},
  year = {2020},
  month = jan 
}

@article{Moreira2010,
  title = {Advances and challenges in explaining fuel spray impingement: How much of single droplet impact research is useful?},
  volume = {36},
  ISSN = {0360-1285},
  url = {http://dx.doi.org/10.1016/j.pecs.2010.01.002},
  DOI = {10.1016/j.pecs.2010.01.002},
  number = {5},
  journal = {Progress in Energy and Combustion Science},
  publisher = {Elsevier BV},
  author = {Moreira,  A.L.N. and Moita,  A.S. and Panão,  M.R.},
  year = {2010},
  month = oct,
  pages = {554–580}
}

@article{Breitenbach2018,
  title = {From drop impact physics to spray cooling models: a critical review},
  volume = {59},
  ISSN = {1432-1114},
  url = {http://dx.doi.org/10.1007/s00348-018-2514-3},
  DOI = {10.1007/s00348-018-2514-3},
  number = {3},
  journal = {Experiments in Fluids},
  publisher = {Springer Science and Business Media LLC},
  author = {Breitenbach,  Jan and Roisman,  Ilia V. and Tropea,  Cameron},
  year = {2018},
  month = feb 
}

@article{Liang2016,
  title = {Review of mass and momentum interactions during drop impact on a liquid film},
  volume = {101},
  ISSN = {0017-9310},
  url = {http://dx.doi.org/10.1016/j.ijheatmasstransfer.2016.05.062},
  DOI = {10.1016/j.ijheatmasstransfer.2016.05.062},
  journal = {International Journal of Heat and Mass Transfer},
  publisher = {Elsevier BV},
  author = {Liang,  Gangtao and Mudawar,  Issam},
  year = {2016},
  month = oct,
  pages = {577–599}
}

@book{Collision2017,
  author    = {Yarin, Alexander L. and Roisman, Ilia V. and Tropea, Cameron},
  title     = {Collision Phenomena in Liquids and Solids},
  publisher = {Cambridge University Press},
  address   = {Cambridge},
  year      = {2017}
}

@article{Cheng2022,
  title = {Drop Impact Dynamics: Impact Force and Stress Distributions},
  volume = {54},
  ISSN = {1545-4479},
  url = {http://dx.doi.org/10.1146/annurev-fluid-030321-103941},
  DOI = {10.1146/annurev-fluid-030321-103941},
  number = {1},
  journal = {Annual Review of Fluid Mechanics},
  publisher = {Annual Reviews},
  author = {Cheng,  Xiang and Sun,  Ting-Pi and Gordillo,  Leonardo},
  year = {2022},
  month = jan,
  pages = {57–81}
}

@article{CLANET2004,
  title = {Maximal deformation of an impacting drop},
  volume = {517},
  ISSN = {1469-7645},
  url = {http://dx.doi.org/10.1017/s0022112004000904},
  DOI = {10.1017/s0022112004000904},
  journal = {Journal of Fluid Mechanics},
  publisher = {Cambridge University Press (CUP)},
  author = {Clanet, Christophe and B{\'e}guin, C{\'e}dric and Richard, Denis and Qu{\'e}r{\'e}, David},
  year = {2004},
  month = sep,
  pages = {199--208}
}

@article{Sanjay2025,
  title = {Unifying Theory of Scaling in Drop Impact: Forces and Maximum Spreading Diameter},
  volume = {134},
  ISSN = {1079-7114},
  url = {http://dx.doi.org/10.1103/PhysRevLett.134.104003},
  DOI = {10.1103/physrevlett.134.104003},
  number = {10},
  journal = {Physical Review Letters},
  publisher = {American Physical Society (APS)},
  author = {Sanjay,  Vatsal and Lohse,  Detlef},
  year = {2025},
  month = mar 
}

@article{Wildeman2016,
  title = {On the spreading of impacting drops},
  volume = {805},
  ISSN = {1469-7645},
  url = {http://dx.doi.org/10.1017/jfm.2016.584},
  DOI = {10.1017/jfm.2016.584},
  journal = {Journal of Fluid Mechanics},
  publisher = {Cambridge University Press (CUP)},
  author = {Wildeman,  Sander and Visser,  Claas Willem and Sun,  Chao and Lohse,  Detlef},
  year = {2016},
  month = sep,
  pages = {636–655}
}

@article{Josserand2016,
  title = {Drop Impact on a Solid Surface},
  volume = {48},
  ISSN = {1545-4479},
  url = {http://dx.doi.org/10.1146/annurev-fluid-122414-034401},
  DOI = {10.1146/annurev-fluid-122414-034401},
  number = {1},
  journal = {Annual Review of Fluid Mechanics},
  publisher = {Annual Reviews},
  author = {Josserand,  C. and Thoroddsen,  S.T.},
  year = {2016},
  month = jan,
  pages = {365–391}
}

@article{Yarin2006,
  title = {DROP IMPACT DYNAMICS: Splashing,  Spreading,  Receding,  Bouncing…},
  volume = {38},
  ISSN = {1545-4479},
  url = {http://dx.doi.org/10.1146/annurev.fluid.38.050304.092144},
  DOI = {10.1146/annurev.fluid.38.050304.092144},
  number = {1},
  journal = {Annual Review of Fluid Mechanics},
  publisher = {Annual Reviews},
  author = {Yarin,  A.L.},
  year = {2006},
  month = jan,
  pages = {159–192}
}

@misc{Zhang2026,
  doi = {10.48550/ARXIV.2601.19835},
  url = {https://arxiv.org/abs/2601.19835},
  author = {Zhang,  Ziyao and Castrejon-Pita,  Alfonso A. and Mostert,  Wouter},
  title = {Numerical simulations of simultaneous pair-drop impacts and their energetics},
  publisher = {arXiv},
  year = {2026},
  copyright = {arXiv.org perpetual,  non-exclusive license}
}

@article{Riboux2015,
  title = {The diameters and velocities of the droplets ejected after splashing},
  volume = {772},
  ISSN = {1469-7645},
  url = {http://dx.doi.org/10.1017/jfm.2015.223},
  DOI = {10.1017/jfm.2015.223},
  journal = {Journal of Fluid Mechanics},
  publisher = {Cambridge University Press (CUP)},
  author = {Riboux,  Guillaume and Gordillo,  José Manuel},
  year = {2015},
  month = may,
  pages = {630–648}
}

@article{Riboux2016,
  title = {Maximum drop radius and critical Weber number for splashing in the dynamical Leidenfrost regime},
  volume = {803},
  ISSN = {1469-7645},
  url = {http://dx.doi.org/10.1017/jfm.2016.496},
  DOI = {10.1017/jfm.2016.496},
  journal = {Journal of Fluid Mechanics},
  publisher = {Cambridge University Press (CUP)},
  author = {Riboux,  Guillaume and Gordillo,  José Manuel},
  year = {2016},
  month = aug,
  pages = {516–527}
}

@article{Gordillo2019,
  title = {A theory on the spreading of impacting droplets},
  volume = {866},
  ISSN = {1469-7645},
  url = {http://dx.doi.org/10.1017/jfm.2019.117},
  DOI = {10.1017/jfm.2019.117},
  journal = {Journal of Fluid Mechanics},
  publisher = {Cambridge University Press (CUP)},
  author = {Gordillo,  José Manuel and Riboux,  Guillaume and Quintero,  Enrique S.},
  year = {2019},
  month = mar,
  pages = {298–315}
}

@article{Roisman2009,
  title = {Inertia dominated drop collisions. II. An analytical solution of the Navier–Stokes equations for a spreading viscous film},
  volume = {21},
  ISSN = {1089-7666},
  url = {http://dx.doi.org/10.1063/1.3129283},
  DOI = {10.1063/1.3129283},
  number = {5},
  journal = {Physics of Fluids},
  publisher = {AIP Publishing},
  author = {Roisman,  Ilia V.},
  year = {2009},
  month = may 
}

@article{Eggers2010,
  title = {Drop dynamics after impact on a solid wall: Theory and simulations},
  volume = {22},
  ISSN = {1089-7666},
  url = {http://dx.doi.org/10.1063/1.3432498},
  DOI = {10.1063/1.3432498},
  number = {6},
  journal = {Physics of Fluids},
  publisher = {AIP Publishing},
  author = {Eggers,  Jens and Fontelos,  Marco A. and Josserand,  Christophe and Zaleski,  Stéphane},
  year = {2010},
  month = jun 
}

@article{Goswami2023,
  title = {Simultaneous impact of droplet pairs on solid surfaces},
  volume = {961},
  ISSN = {1469-7645},
  url = {http://dx.doi.org/10.1017/jfm.2023.249},
  DOI = {10.1017/jfm.2023.249},
  journal = {Journal of Fluid Mechanics},
  publisher = {Cambridge University Press (CUP)},
  author = {Goswami,  Anjan and Hardalupas,  Yannis},
  year = {2023},
  month = apr 
}

@article{Riboux2014,
  title = {Experiments of Drops Impacting a Smooth Solid Surface: A Model of the Critical Impact Speed for Drop Splashing},
  volume = {113},
  ISSN = {1079-7114},
  url = {http://dx.doi.org/10.1103/PhysRevLett.113.024507},
  DOI = {10.1103/physrevlett.113.024507},
  number = {2},
  journal = {Physical Review Letters},
  publisher = {American Physical Society (APS)},
  author = {Riboux,  Guillaume and Gordillo,  José Manuel},
  year = {2014},
  month = jul 
}

\end{document}